\documentclass[12pt,preprint]{emulateapj}
\usepackage{graphicx}

\shorttitle{KOI-1003:  A new spotted, eclipsing RS CVn}

\shortauthors{Roettenbacher et al.}
\slugcomment{Accepted for publication in ApJ}

\begin{document}

\title{KOI-1003: A new spotted, eclipsing RS CVn binary in the \emph{Kepler} field}

\author{
  Rachael M.\ Roettenbacher$^{1,2}$,
  Stephen R.\ Kane$^3$,
  John D.\ Monnier$^1$, and
  Robert O.\ Harmon$^4$
  }
  \affil{$^1$Department of Astronomy, University of Michigan, Ann Arbor, MI 48109, USA \\  
  $^2$Department of Astronomy, Stockholm University, SE-106 91 Stockholm, Sweden \\
  $^3$Department of Physics and Astronomy, San Francisco State University, San Francisco, CA 94132, USA \\ 
  $^4$Department of Physics and Astronomy, Ohio Wesleyan University, Delaware, OH 43015, USA \\ 
  }
\email{Contact email:  rachael@astro.su.se}


\begin{abstract}

Using the high-precision photometry from the \emph{Kepler} space telescope, thousands of stars with stellar and planetary companions have been observed.  The characterization of stars with companions is not always straightforward and can be contaminated by systematic and stellar influences on the light curves.  Here, through a detailed analysis of starspots and eclipses, we identify KOI-1003 as a new, active RS CVn star---the first identified with data from \emph{Kepler}.  The \emph{Kepler} light curve of this close binary system exhibits the system's primary transit, secondary eclipse, and starspot evolution of two persistent active longitudes.  The near equality of the system's orbital and rotation periods indicates the orbit and primary star's rotation are nearly synchronized ($P_\mathrm{orb} = 8.360613\pm0.000003$~days; $P_\mathrm{rot} \sim 8.23$~days).  By assuming the secondary star is on the main sequence, we suggest the system consists of a $1.45^{+0.11}_{-0.19} \ M_\odot$ subgiant primary and a $0.59^{+0.03}_{-0.04} \ M_\odot$ main-sequence companion.  Our work gives a distance of $4400 \pm 600$~pc and an age of $t = 3.0^{-0.5}_{+2.0}$~Gyr, parameters which are discrepant with previous studies that included the star as a member of the open cluster NGC 6791.  

\end{abstract}

\keywords{binaries:  eclipsing -- stars: activity -- stars:  imaging -- stars:  variables:  general -- starspots --  stars: individual (KOI-1003)}


\section{Introduction}
\label{intro}

The high-precision, nearly-continuous photometry obtained by the \emph{Kepler} satellite overcame the limitations of ground-based photometry to allow for unprecedented studies of $>10^5$ stars.  Working toward the primary goal of the mission, the number of known and candidate exoplanets has dramatically increased through the analysis of \emph{Kepler} photometry \citep[e.g.,][]{bor11a, bor11b,bat13}, but the photometry has also provided a wealth of information for stellar astrophysics, including asteroseismology, stellar properties, and stellar activity \citep[e.g.,][]{gil10,hub13,roe13}.  

The activity of stars with convective envelopes includes starspots---dark, cool regions of the photosphere that are caused by magnetic fields stifling the convection in the outer layers of the stars \citep[e.g.,][]{str09}.  Using \emph{Kepler} photometry, starspots have been studied with spot modeling \citep[e.g.,][]{fra11, fro12} and light-curve inversion \citep[e.g.,][]{sav11, roe13} techniques to produce surface maps of spot structures that show spot evolution and differential rotation.  

A particularly interesting class of spotted, active stars are RS Canum Venaticorum (RS CVn) stars.  RS CVns are often close binary systems with an evolved, partially Roche-potential-filling giant or subgiant primary star and a much fainter, less-evolved subgiant or dwarf secondary star \citep{ber05, str09}.  The systems are photometrically variable and show evidence of fluctuations in Ca H and K \citep{hal76}.  As binaries, the detection of the faint companion allows for an understanding of the evolutionary history of the system, including stellar parameter estimates \citep[e.g.,][]{roe15a,roe15b}.  Multiple-epoch studies of RS CVn stars have allowed for measures of differential rotation and spot evolution \citep[e.g.][]{kov12,roe11}.  An improved understanding of the system and spot parameters for these stars will shed light on analogous systems (e.g., T Tauri stars).  

In the \emph{Kepler} Input Catalog (KIC), many stars exhibit variable light curves \citep{bas10, bas11, bas13}, a number of which are likely the result of rotational modulation due to starspots.  Because the starspots of RS CVns evolve \citep[e.g.,][]{hen95,roe11}, and they are typically binary systems, we focus our study on \emph{Kepler} stars that show evidence of stable, yet evolving starspots and eclipses.  KOI-1003 (KIC 2438502, 2MASS J19211869+3743362) is one such star with a rotationally-modulated light curve attributed to starspots and their evolution, as well as exhibiting a primary transit and secondary eclipse.  

In this detailed study of an RS CVn system using the \emph{Kepler} satellite light curves, we study both the starspots and eclipses of KOI-1003.  With the precision of the \emph{Kepler} light curve and models of the system's eclipses, we are better able to classify the classic RS CVn system through constraining orbital parameters.  This allows us to improve upon previous estimates, such as rejecting the star's membership of the old ($\sim 8$ Gyr) open cluster NGC 6791 by past studies \citep{moc02,moc05}.  These studies also used ground-based photometry to refine a photometric period to 8.3141~days but could not establish the system as eclipsing, a characteristic that is unmistakeable in the \emph{Kepler} light curve.  

In this paper, we present an analysis of the light curve of KOI-1003, which includes a discussion of the \emph{Kepler} characterization of the star as a false positive and our work to more accurately characterize the system components through analysis of the transit and eclipse.  We also present surface reconstructions of the primary star's starspots using a light-curve inversion algorithm.   
In Section 2, we present the \emph{Kepler} observations and known parameters of KOI-1003.  In Section 3, we discuss the disposition of the eclipsing binary KOI-1003, transit depths, the Keplerian orbital parameters of the system, and new estimates of stellar parameters.  We discuss the spot models and persistent spots in Section 4.  We conclude in Section 5 with a discussion of our findings.  We include an appendix of surface reconstructions of all 164 rotational epochs observed by \emph{Kepler}.


\section{Observations}

\begin{figure*}
  \includegraphics[angle=0,scale=.85]{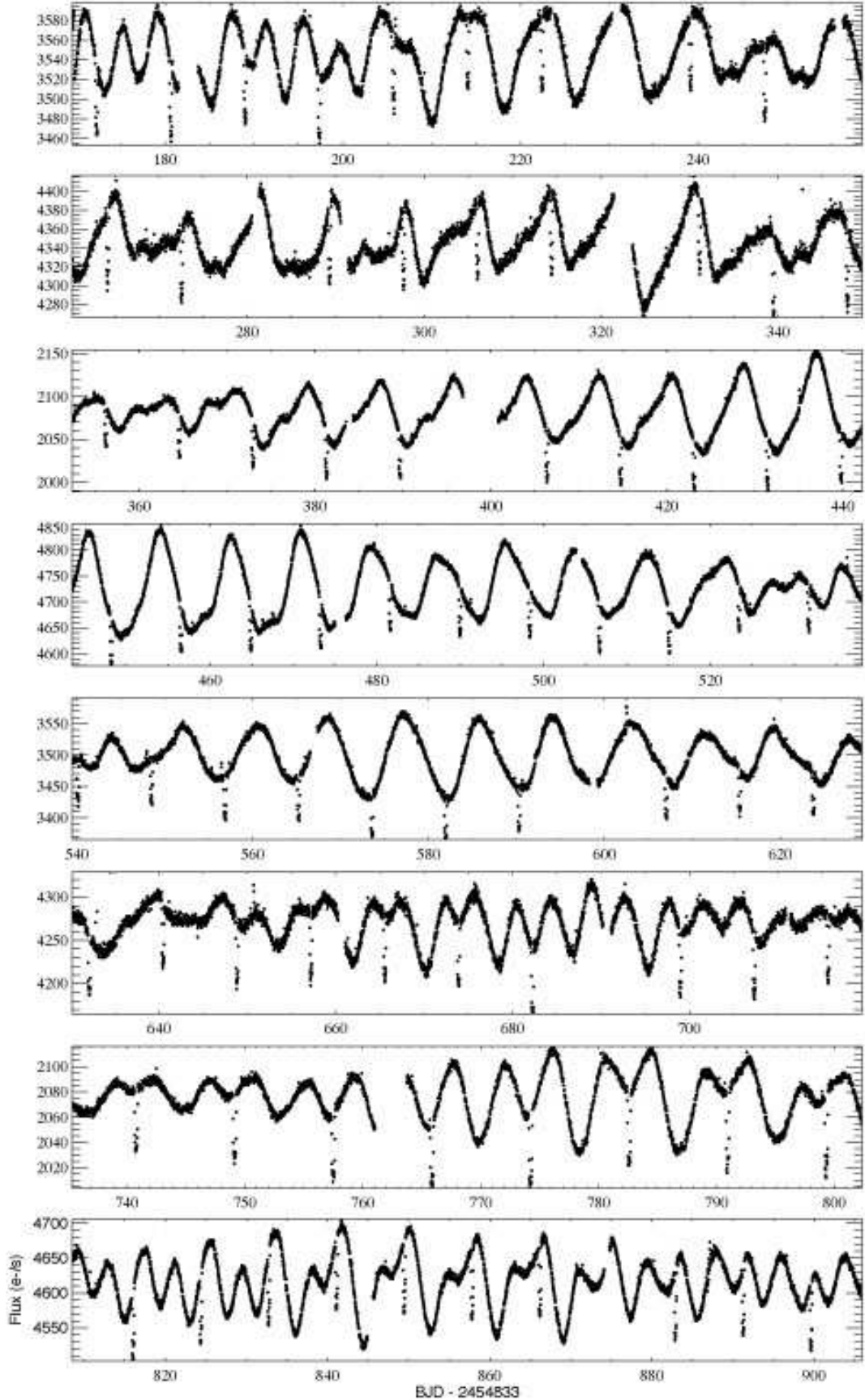}
  \caption{Long cadence \emph{Kepler} SAP light curves of KOI-1003 for Q2-9 (top panel is Q2 with the quarter increasing down the page) with CBVs removed.  The panels are scaled to each quarter.  For more details, see Section 2.
  \label{fulllc1}}
\end{figure*}

\begin{figure*}
  \includegraphics[angle=0,scale=.85]{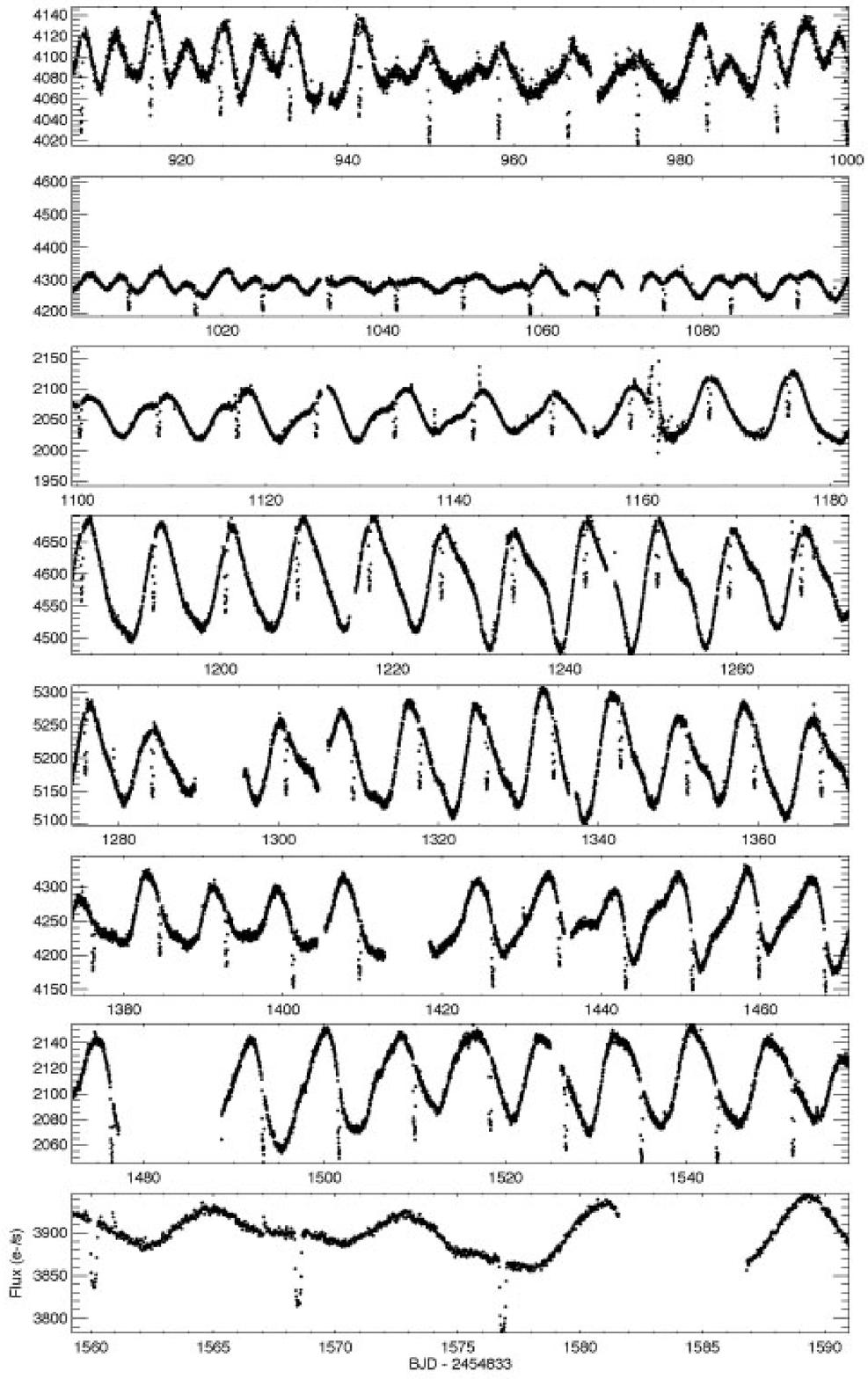}
  \caption{Long cadence \emph{Kepler} SAP light curves of KOI-1003 for Q10-17.  The plots are as in Figure \ref{fulllc1}.  
  \label{fulllc2}}
\end{figure*}

KOI-1003 \citep[$K_p =16.209$,][]{bro11} was observed by the \emph{Kepler} space telescope \citep{bor10,koc10} nearly continuously with long-cadence (29.4 min) observations in \emph{Kepler} Quarters 2-17 as a target of exoplanet and Guest Observer programs.  In our analysis of the system's eclipses, we use the pre-search data conditioning (PDC) light curve.  For our analysis of the primary star's surface features, we use the \emph{Kepler} simple aperture photometry (SAP) light curve with the  cotrending basis vectors (CBVs) removed, as this method best conserves the stellar astrophysics in the light curve.  We remove the CBVs from the SAP light curve using the \emph{kepcotrend} tool of the PyKE software package \citep[][see Figures \ref{fulllc1} and \ref{fulllc2}]{sti12}.  The CBVs used depended upon the quarter (two CBVs used in Q4, 8, 10, 11, 12, 13, 15, and 17; three used in Q3, 5, 6, 7, 9, 14, and 16; and five used in Q2).  After removing the systematic effects, the remaining signal in the SAP/CBV light curve is assumed to be that of the eclipse, white-light flares, and starspots.  

\begin{figure}
\includegraphics[angle=90,scale=.35]{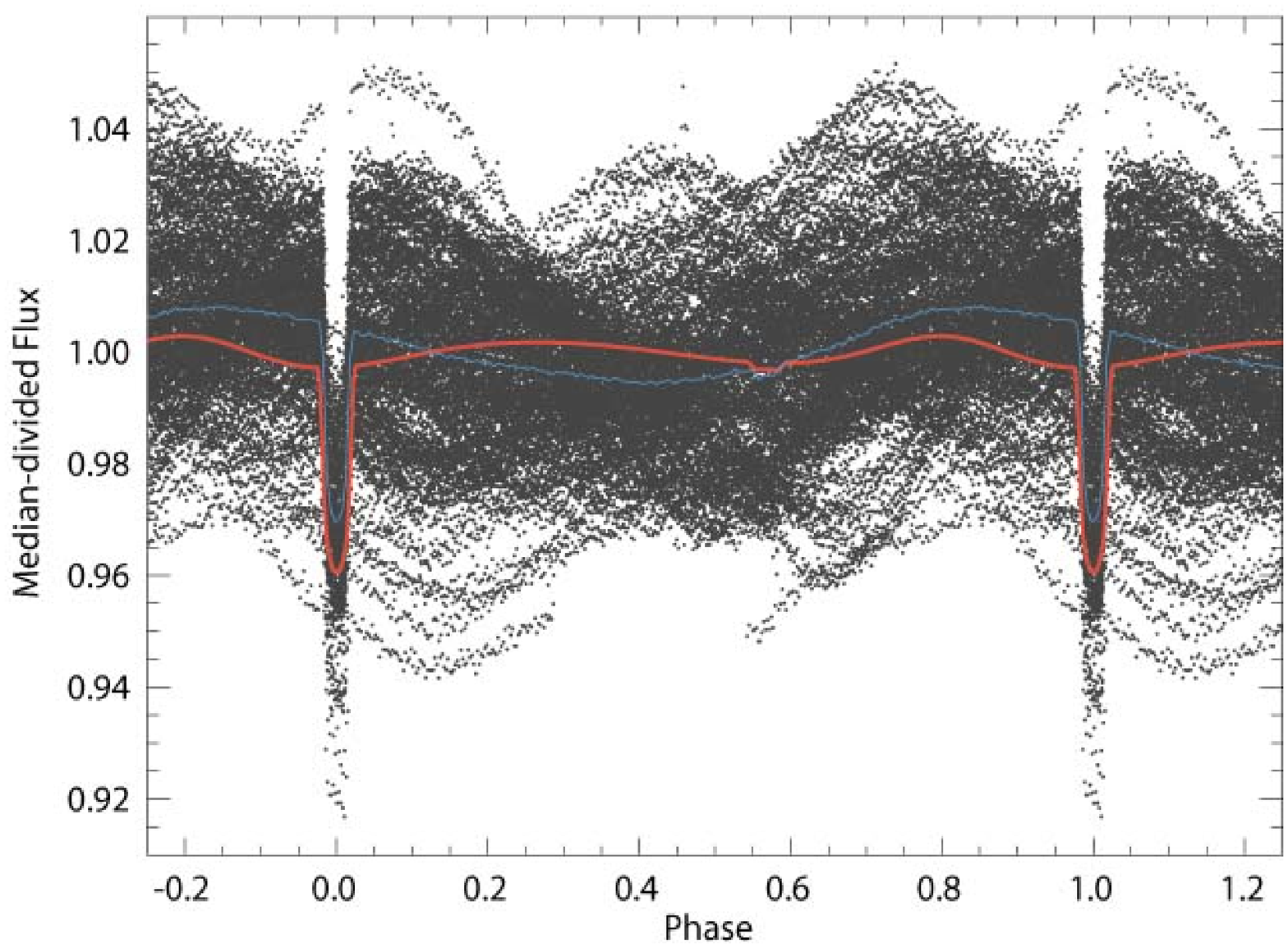}
\caption{Folded, CBV-corrected SAP \emph{Kepler} light curves of KOI-1003.  Each data point is a long-cadence measurement from the \emph{Kepler} light curve (Q2-17).  A primary transit is observed at phase $0.000000 \pm 0.000010$ with a secondary eclipse at phase $0.5695 \pm 0.0005$.   Overlaid as the solid blue line is the folded and binned (150 bins in phase) CBV-corrected SAP \emph{Kepler} light curves of KOI-1003.  We do not include standard deviation error bars as these are dominated by fluctuations caused by the starspots and $P_\mathrm{orb} \ne P_\mathrm{rot}$.  Note the presence of the starspot residual (quasi-sinusoidal feature minimum at $\phi \sim 0.4$) and the amplitudes of the transit and eclipse are reduced due to the presence of transient starspots.  The (thicker) solid red line is modeled in ELC \citep{oro00} using the observed and estimated parameters of the primary and secondary components of KOI-1003 (see below).  The ELC-modeled light curve shows evidence of ellipsoidal variations, which contribute to the light curve nearly two orders of magnitude less than many of the starspots present on the stellar surface during observations by \emph{Kepler}.
  \label{KIC2438502fold}}
\end{figure}

For the complete CBV-removed \emph{Kepler} light curve (Q2-17) we stitch the light curves together by a simple median-division.  We folded the data over the NASA Exoplanet Archive's orbital period ($P_\mathrm{orb} = 8.360613 \pm 0.000003$ days; transit epoch (BJD $-$ 2454833), $T = 172.2653 \pm 0.0003$)  averaging within 150 phase bins (see Figure \ref{KIC2438502fold}).  The large-amplitude quasi-sinusoidal modulation is due to starspots.   We do not include error bars derived from the standard deviation of the fluxes in the individual bins, as they have a large value due to the evolving spot structures and differential rotation on the surface.

The prominent transit that triggered \emph{Kepler} Object of Interest classification of the system is located at phase $0.00000 \pm 0.00001$.  The secondary eclipse is located at phase $0.5695 \pm 0.0005$, which is detected in the \emph{Kepler} Data Validation Report (DVR) for quarters 1-17\footnote{\tt exoplanetarchive.ipac.caltech.edu/data/KeplerData/002/\\ 002438/002438502/dv/kplr002438502-20141002224145\_dvr.pdf}, but is labeled as ``Planet 2.''  Because this eclipse does not occur exactly half of an orbit from the primary eclipse, the orbit must be eccentric.  We estimate the system's eccentricity to be $e = 0.113 \pm 0.012$ ($\omega = 346.^\circ \pm 20.^\circ$; see Section 3.2).


\section{The Nature of KOI-1003}
\label{disposition}

The disposition of KOI-1003 has changed several times over the course
of the \emph{Kepler} observations.  In \citet{bor11b}, the star was first listed as a \emph{Kepler} candidate in the Q0-2 data release. The object retained a disposition of ``candidate'' in
the Q1-6 data release \citep{bat13}, but 
\citet{bur14} changed the disposition of the object to ``not
dispositioned.''  According to the NASA Exoplanet Archive\footnote{\tt http://exoplanetarchive.ipac.caltech.edu/},
the ``cumulative'' \emph{Kepler}
data release modified the disposition to ``false positive'' and the
subsequent Q1-16 data release changed the status back to ``not
dispositioned.''  After analysis of the complete light curve, the system is currently listed as a false positive.  

The Q1-17 DVR for this system
indicates that there are two major causes of the disposition
discrepancies:  the presence of a secondary eclipse and an
apparent offset of the PSF centroid compared with out-of-transit
observations. These centroid offsets are generally inside the
3-$\sigma$ radius of confusion for the weighted mean offset, with the
exception of quarters Q5, 9, 13, and 17.  As the \emph{Kepler} spacecraft
rotated $90^\circ$ every 90 days, these four discrepant quarters occurred separated in time by a full \emph{Kepler} rotation such that star fell on the same pixels. 
Examination of the pixel mask used for these quarters shows
that there are no detected nearby stars that caused the
significant centroid offsets described in the DVR. The fit
location of the Pixel Response Function (PRF) always fell in the same
pixel for these anomalous quarters (Column 149, Row 925, Module 10, Channel 29). This pixel is
not listed in the pre-launch bad pixel map (Douglas Caldwell; private
communication); however, this list is known to be incomplete.

The presence of nearby stars can often be an additional source of confusion for \emph{Kepler} due to the relatively large pixel size ($3.98 \arcsec \times 3.98 \arcsec$). As such, high-spatial-resolution imaging of the field surrounding \emph{Kepler} candidates forms a major component of \emph{Kepler} follow-up activities \citep[e.g.,][]{ada12,ada13}.  The Target Pixel Files for KOI-1003 show that the signature of both the starspots and eclipses are related to the flux of KOI-1003.  However, the \emph{Kepler} DVR shows that the out-of-transit centroid to be shifted such that the star itself does not fall in the radius of confusion suggesting flux from nearby stars may be contaminating the KOI-1003 light curve.  

\begin{figure}
\begin{centering}
  \includegraphics[angle=90,scale=.3]{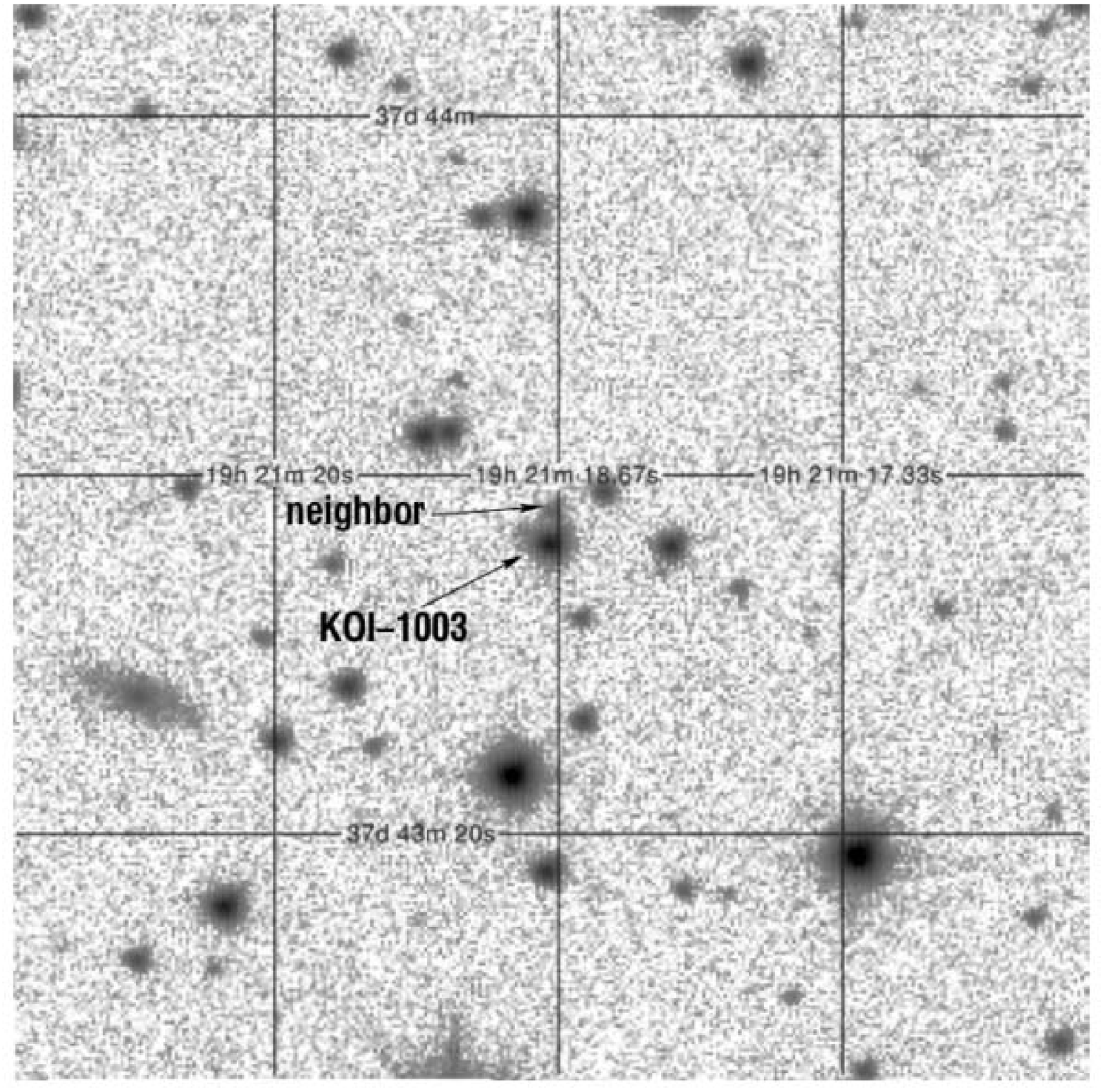}
  \caption{$J$-band UKIRT image of the KOI-1003 field \citep[$1\arcmin \times 1 \arcmin$; $J = 14.68$;][]{cut03}. The
    star KOI-1003 is labeled in the center of the frame. A potential
    source of contamination is the neighboring star directly above
    KOI-1003.  The neighbor contributes $\sim 2\%$ of the ($J$-band) flux to the system, which is too small to account for the transits or starspots present in the light curve.  The flux from the neighbor is neglected in our analysis.    }
  \label{ukirt}
  \end{centering}
\end{figure}

While there are no bright stars within a pixel of KOI-1003, there is a faint neighboring star that is visible in the $J$-band image including KOI-1003 from the United Kingdom Infrared Telescope (UKIRT) survey\footnote{\tt http://www.ukirt.hawaii.edu/} (see Figure \ref{ukirt}) but is unresolved by \emph{Kepler}.  From the UKIRT data, simple aperture photometry of the stars shows that the flux received from the neighboring star is $\sim 2$\% ($J$-band) of that received from KOI-1003.  The additional flux of this star in combination with that of other nearby stars will not significantly dilute the transit depth.

\subsection{Variable Transit Depths}
\label{depthsec}

\begin{figure}
  \includegraphics[angle=90,scale=.35]{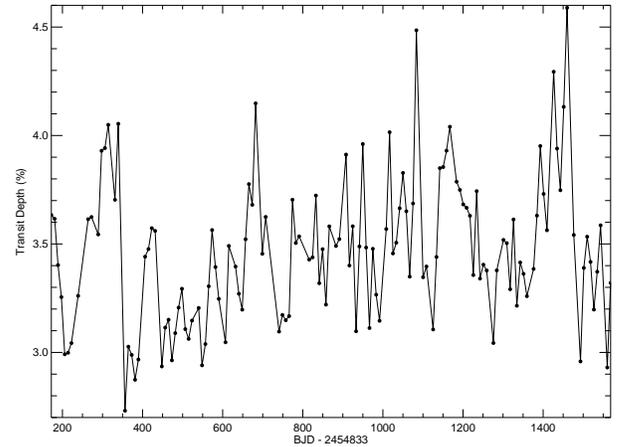}
  \caption{The measured primary transit depth for the KOI-1003 system
    as a function of time.  The depth of the eclipse varies as measured in the PDC light curve due to the presence of starspots in the path of the transit. } 
      \label{depthfig}
\end{figure}

A side effect of stellar activity could be that the
amount of light blocked during a transit changes
as a function of time if the transit crosses the starspots \citep[e.g.,][]{sil10,san11}.  Variations in the transit depth are observed
for the KOI-1003 system during the course of \emph{Kepler}
observations.  Using the PDC light curve, we assumed a linear trend representing features across the transit.  We then simply averaged the flux before and after the transit and divided that value from the flux to eliminate the large-scale influence of starspots across the transit.  Dividing this corrected light curve by the mean of the out-of-transit measurements, we found the depth of the primary transit.
By this method, we found that the individual transit depths vary between $2.73\%$ and $4.59\%$ of flux with a mean of $3.46\pm0.33\%$ (see Figure \ref{depthfig}).  

\begin{figure}
  \includegraphics[angle=90,scale=.35]{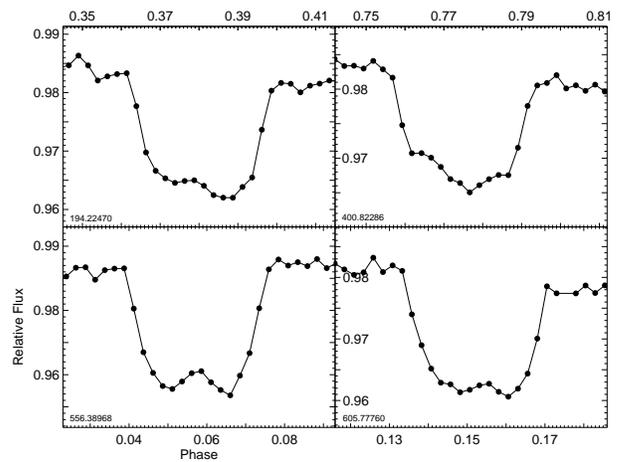} 
  \caption{Sample of four transit spot-crossing events (unbinned, PDC light curves).  The small increases in flux during the transit are due to the path of the transit crossing over starspots. The start date of each rotation period is recorded in the lower left of the panel (Barycentric Julian Date $-$ $2454833$).}
  \label{spotsineclipse}
\end{figure}

We suggest that the transit depth variations are due, at least in part, to the companion transiting the spotted surface.  Should the companion cross a starspot as it transits, an increase in flux will be observed in the transit, changing its depth \citep[e.g., ][]{san11, cro15}.  We observe this in a number of eclipses when starspots are present on the face of the star that the transit crosses.  In Figure \ref{spotsineclipse}, we present four examples of spot-crossing events.  

In order to better quantify the depth of the transit and determine further system parameters, we employed the stellar and orbital parameter fitting software EXOFAST 
\citep{eas13} to determine a primary transit depth of $\delta_\mathrm{t} = 3.14 \pm 0.10 \%$, consistent within 1-$\sigma$ errors with our simplistic result above.  A detailed description of these efforts is found in the next subsection.

\subsection{Keplerian Orbital Elements}
\label{orbelem}

The Keplerian orbital elements of eccentricity, $e$, and argument of periastron, $\omega$, 
may be determined for a companion when there is a secondary eclipse
detected in the photometric time series \citep{cha05}. 
In order to estimate the timing and duration of the transit and eclipse, as well as $e$ and $\omega$, we use EXOFAST, a suite of routines that uses a differential evolution Markov chain Monte Carlo method to determine orbital and stellar parameters using transit and/or radial velocity data \citep{eas13}.

Before running EXOFAST on KOI-1003, we removed the signature of the starspots by isolating the times of transit and eclipse and by fitting a line across the data just before and after each transit and eclipse.  We removed that linear trend and set all out-of-transit and out-of-eclipse data points to a normalized flux of 1.0.  We folded and binned the complete transit and eclipse data set across all \emph{Kepler} quarters to obtain the best-fit values for KOI-1003.  We additionally broke the spotless light curve into three independent chunks of data, in order to ascertain the error on the orbital parameters (the standard deviation of which is assigned as the 1-$\sigma$ error for the parameters).  These independent chunks of data were similarly folded and binned.  Each folded and binned light curve was then run through EXOFAST fixing the orbital period, $P_\mathrm{orb}$; time of central transit, $T_\mathrm{C}$; and primary star effective temperature $T_\mathrm{eff,1}$.  We placed weak constraints on metallicity, [Fe/H], and surface gravity, log $g$.

Using EXOFAST on the folded and binned PDC light curves, we determined
the timing of primary transit and secondary eclipse (in phase units)
to be $t_\mathrm{t} = 0.00000 \pm 0.00001$ and $t_\mathrm{e} = 0.5695 \pm 0.0005$,
respectively.
Using the equations of Keplerian orbits, EXOFAST provides $e = 0.113 \pm 0.012$ and $\omega = 346.^\circ \pm 20.^\circ$.  The total duration of the transit and eclipse are (in phase units) $0.0407 \pm 0.0002$ and $0.039 \pm 0.003$, respectively.  The depth of the transit and eclipse are $3.14 \pm 0.10\%$ and $0.176 \pm 0.012\%$, respectively.
EXOFAST used the eclipse depths to determine the ratio of the radii of the secondary to the primary, $0.177 \pm 0.003$, and the ratio of the primary radius to the semi-major axis, $8.2 \pm 0.5$. For the complete set of parameters, see Table \ref{EXOFASTparam}.

\begin{deluxetable}{l c}
\tabletypesize{\scriptsize}
\tablewidth{0pt}
  \tablecaption{\label{EXOFASTparam} System Parameters for KOI-1003 from EXOFAST}
  \tablehead{
    \colhead{Parameter } &
    \colhead{Value}
  }
  \startdata
Eccentricity, $e$ & $0.113 \pm 0.012$ \\
Argument of periastron, $\omega$ ($^\circ$) & $346. \pm 20.$ \\
Orbital angle of inclination, $i_\mathrm{orb}$ ($^\circ$) & $86.0 \pm 0.5$ \\
Ratio of secondary to primary radii ($R_2/R_1$) & $0.177 \pm 0.003$ \\
Ratio of semi-major axis to primary radii ($a/R_1$) & $8.2 \pm 0.5$\\
Timing of primary transit, $t_\mathrm{t}$ (phase units) & $ 0.000000 \pm 0.000010$  \\
Duration of primary transit, $T_\mathrm{t}$ (phase units) & $0.0407 \pm 0.0002$\\
Depth of primary transit, $\delta_\mathrm{t}$ ($\%$) & $3.14 \pm 0.10$ \\
Primary transit impact parameter, $b_\mathrm{t}$ & $0.582 \pm 0.005$ \\
Timing of secondary eclipse, $t_\mathrm{e}$ (phase units) & $0.5695 \pm 0.0005$\\
Duration of secondary eclipse, $T_\mathrm{e}$ (phase units) & $0.039 \pm 0.003$ \\
Depth of secondary eclipse, $\delta_\mathrm{e}$ ($\%$) & $0.176 \pm 0.012$\\
Secondary eclipse impact parameter, $b_\mathrm{e}$ & $0.55 \pm 0.07$
  \enddata
\end{deluxetable}



\subsection{Stellar Parameters}
\label{priparam}

KOI-1003 is listed in the \emph{Kepler} Input Catalog as an early K-type star with temperature $T_\mathrm{eff,1} \sim 5200$ K, $\log g_1 \sim 4.5$, and metallicity [Fe/H] $\sim -0.1$ \citep[NASA Exoplanet Archive;][]{bro11}\footnote{The stellar parameters given by the \emph{Kepler} Input Catalog have been shown to sometimes be inaccurate; however, in the case of KOI-1003, the results from the \emph{Kepler} Input Catalog, \citet{pin12}, and \citet{hub14} are all consistent.}.  \citet{moc02} name KOI-1003 as a member of NGC 6791 (labeling the star as NGC 6791 KR V54).  KOI-1003 is located outside of the core of the cluster and is identified as a member of the cluster based only on proximity, as velocities to confirm membership have not been obtained.  \citet{cha99} give NGC 6791 an age of $\sim8$ Gyr.  

With these prior determinations in mind, we investigated the nature of the primary star through an analysis using the orbital elements and parameters determined in Section \ref{orbelem} by EXOFAST.
From the radius ratio from EXOFAST of $0.177 \pm 0.003$, we suggest that the primary star is not a main sequence star as described in the NASA Exoplanet Archive.  We suggest that the primary star is a subgiant, which is consistent with latest catalog released by the \emph{Kepler} Stellar Properties Working Group \citep{mat16}, and the companion is a sub-solar-mass main sequence star.

\begin{figure}
  \includegraphics[angle=90,scale=.35]{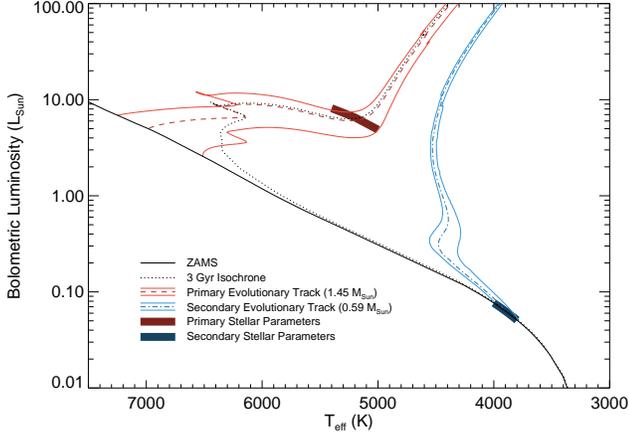} 
  \caption{H-R diagram for KOI-1003.  The red, dashed and blue, dot-dashed lines are the solar metallicity, main and post-main sequence evolutionary tracks for $1.45 \ M_\odot$ and $0.59 \ M_\odot$ stars, respectively.  These values assume $T_\mathrm{eff,1} = 5200$~K.  The solid lines above surrounding this best-fit model are the models that result when $T_\mathrm{eff,1} = 5000$ and 5400~K, represented by the cooler and hotter tracks, respectively.  The solid black line the zero age main sequence, and the dotted line is a 3 Gyr isochrone \citep{dot08}.  The dark red box crossing the more massive tracks is the predicted location the primary star of KOI-1003.  The dark blue box along the lower end of the main sequence is the predicted location of the secondary star.}
  \label{KIC2438502HR}
\end{figure}

Assuming that the secondary star is on the main sequence, we used the Dartmouth stellar evolution models \citep{dot08} to restrict the masses and radii for a range of temperatures of the secondary star.  We then used the radius ratio from EXOFAST and the assumed primary temperature of $T_\mathrm{eff,1} = 5200 \pm 200$ K to determine potential masses for the primary star.  We further assumed $\log g_\mathrm{1}$ values appropriate for a subgiant.  We found that $\log g_\mathrm{1} = 3.6$ yielded the most realistic stars for the given temperature and radius range (e.g., neither sub- nor super- luminous for a star with the required radius).  By this assessment, we found that the best fit to the calculations was the primary star with $T_\mathrm{eff,1} = 5200$ K, $R_\mathrm{1} = 3.16 \ R_\odot$ and $M_\mathrm{1} = 1.45 \ M_\odot$ and a secondary star with $T_\mathrm{eff,2} = 3900$ K, $R_\mathrm{2} =  0.56 \ R_\odot$, and $M_\mathrm{2} = 0.59 \ M_\odot$.  We illustrate these results with the Hertzsprung-Russel (H-R) diagram in Figure \ref{KIC2438502HR} and list the parameters in Table \ref{ELCparams}.   The lower and upper limits are the values associated with the same analysis performed with $T_\mathrm{eff,1} = 5000$ and 5400 K, respectively, as lower and upper limits on the primary star's temperature.  While our values for the mass and radius are internally-consistent, they are not consistent with those calculated by \citet{mat16}.  We note that their mass and radius values for KOI-1003 are not consistent with their reported $\log g$.

\begin{deluxetable}{l c}
\tabletypesize{\scriptsize}
\tablewidth{0pt}
  \tablecaption{\label{ELCparams} Observed and Estimated Stellar Parameters for KOI-1003}
  \tablehead{
    \colhead{ Parameter } &
    \colhead{Estimated Value}
  }
  \startdata
  Primary effective temperature, $T_\mathrm{eff, 1}$ (K) & $5200 \pm 200$ \\
  Primary radius, $R_1$ ($R_\odot$) & $3.2^{+0.2}_{-0.3}$ \\
 Primary luminosity, $L_1$ ($L_\odot$) & $6.6^{+1.6}_{-1.7}$ \\
  Primary mass, $M_1$ ($M_\odot$) & $1.45^{+0.11}_{-0.19}$ \\
  Primary surface gravity, log $g_1$ (cm s$^{-2}$)$^a$ & $3.6$ \\
  Secondary effective temperature, $T_\mathrm{eff, 2}$ (K) & $3900 \pm 100$ \\
  Secondary radius, $R_2$ ($R_\odot$) & $0.56^{+0.02}_{-0.04}$ \\
 Secondary luminosity, $L_2$ ($L_\odot$) & $0.066^{+0.011}_{0.016}$ \\
  Secondary mass, $M_2$ ($M_\odot$) & $0.59^{+0.03}_{-0.04}$ \\
  Secondary surface gravity, log $g_2$ (cm s$^{-2}$) & $5.2^{-0.2}_{+0.4}$\\
  Semi-major axis, $a$ ($R_\odot$) & $22.0^{+0.8}_{-1.5}$ \\
  System metallicity, [Fe/H]$^a$ & 0.0 \\
  System age (Gyr) & $3.0^{-0.5}_{+2.0}$\\
  Limb-darkened diameter, $\theta_\mathrm{LD}$ (mas) & 0.0067 \\
  Distance, $d$ (pc) & $4400 \pm 600$ 
  \enddata
  \tablecomments{Note that we are not precisely determining these parameters, but only estimating them based upon physically reasonable constraints. \\
  $^a$These parameters are assumed.} 
\end{deluxetable}

We verified the temperature of the secondary and the surface gravity estimate by comparing the flux through the \emph{Kepler} bandpass for stars with the properties we determined.  Using NextGen models \citep{hau99}, we used parameters $T_\mathrm{eff,1} = 5200 \pm 200$ K, $\log g_\mathrm{1} = 3.5$, and $\log g_\mathrm{2} = 5.0$.  We found that the flux ratio determined through the EXOFAST fitting was met by requiring $T_\mathrm{eff,2} = 3800 \pm 200$ K, which is consistent with the previously described analysis.  

We do note that this solution is not completely consistent with the EXOFAST-derived values, particularly $a/R_\mathrm{1}$.  This solution would require $a/R_\mathrm{1} \sim 7$, but EXOFAST determined a value of $8.2\pm0.4$.  The best-fit value of $8.22$ is the value achieved when running EXOFAST on the entire data set of transits.  The error bars of $\pm 0.4$ are determined by the splitting the entire data set into three independent sets and running EXOFAST, taking the standard deviation of the best-fit values for those sets.  It is likely that the error we present here is not large enough, as the fits for the individual data sets are $8.4 \pm 0.33$,  $8.3^{+1.0}_{-1.9}$, and $7.7^{+1.1}_{-1.2}$.  By considering these EXOFAST fits, our required $a/R_\mathrm{1} \sim 7$ is less discrepant.  A potential source of this error could be the impact of spot-crossing events (as in Figure \ref{spotsineclipse}) on the eclipse, artificially increasing the uncertainty in the eclipse timing.

With the Dartmouth models, we were also able to assign the system an age of 3 Gyr, significantly younger than the 8 Gyr age suggested by \citet{cha99}.  The difference in age between the cluster NGC 6791 and KOI-1003 and the location of the star away from the core of the cluster suggests that the star is not part of the cluster.  

The distance of the cluster is given by reddening $E(B-V) = 0.10$ and distance modulus $(m-M)_V = 13.42$ \citep{cha99}, which gives a distance of $4170$ pc. In order to determine the distance to KOI-1003 with our analyses, we utilize the Johnson $B$ and $V$ magnitudes given by \citet{moc05}, extinction from the \emph{Kepler} Input Catalog \citep[$E(B-V) = 0.16$][]{bro11}, and the surface brightness relation for color, magnitude, and limb-darkened linear diameter ($\theta_\mathrm{LD}$) from \citet[][their Equation 1]{ker08}.  We find the linear diameter $\theta_\mathrm{LD} = 0.0067$~mas, resulting in a distance of $4400 \pm 600$~pc.  While this is consistent with the distance to NGC 6791, the system and cluster ages are not consistent.  Instead, KOI-1003 is a field RS Canum Venaticorum (RS CVn) system, a class of stars characterized by close, active binaries with an evolved primary and main-sequence companion  \citep[e.g.,][]{ber05,str09}.

With the best-fit parameters determined above (see Tables \ref{EXOFASTparam} and \ref{ELCparams}), we made model light curves of the system using the light-curve-fitting and modeling software package Eclipsing Light Curve \citep[ELC;][]{oro00}.  ELC produces a model light curve for the input parameters (see appropriate line in Figure \ref{KIC2438502fold}).  In this spotless model light curve, we see evidence of the presence of ellipsoidal variations in the system.  Using the \citet{egg83} approximation for the Roche lobe potential, we find that the Roche radius, $R_\mathrm{L} = 12 \ R_\odot$, and the Roche lobe filling is $R_1/R_\mathrm{L} = 0.27$.
The photometric strength of the ellipsoidal variations ($0.2\%$) is significantly weaker than the signature of the starspots (up to $10\%$), thus we do not account for them in our further analysis.



\section{Periodic Light Curve Signatures}

To determine the significant periods present in the
KOI-1003 time series photometry we used a weighted Lomb-Scargle (L-S)
Fourier analysis, similar to that described by \citet{kan07}.  
\citet{aig15} show that auto-correlation function (ACF) period searches are better than periodogram searches at identifying rotational periods for complex starspot structures.  Because the regions of activity on KOI-1003 are well-defined, long-lived, and few in number, either type of period-search would be successful in this case.    To stitch the individual quarter light curves together, we divided the median of each quarter after the CBVs have been removed (see Section 1).

\begin{figure}
  \includegraphics[angle=270,scale=.35]{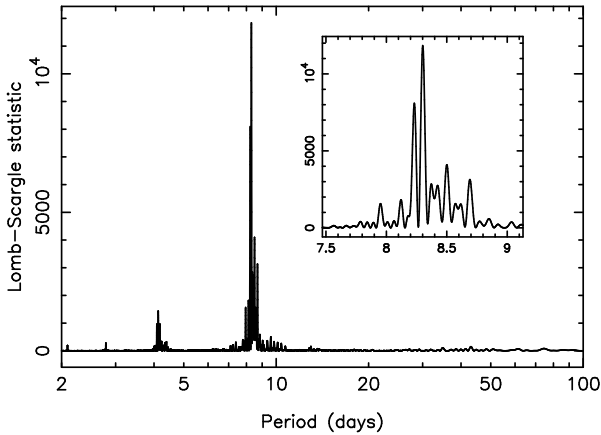}
  \caption{Weighted L-S periodogram of \emph{Kepler} Q2-17 photometry for KOI-1003.  The inset panel shows the detailed Fourier power
    spectrum around the potential rotation period of the primary
    star.}
  \label{pgram}
\end{figure}

The resulting periodogram is shown in Figure \ref{pgram}. The use of
long cadence ($\sim29.4$ minutes) data produced a Nyquist frequency of 24.5 days$^{-1}$ and does not overwhelm the periodogram.
The dominant power in the Fourier
spectrum lies in a region  between $7.5$ and $9$ days (see Figure \ref{pgram} inset) and contains the ten most powerful peaks in  
the periodogram. These peaks likely represent
spot activity at different latitudes over the course of \emph{Kepler}
observations. The orbital period of the companion (8.360613~days) is not
among these periods as the Fourier analysis is optimized toward
detection of sinusoidal rather than transit signatures. We select the
second-strongest period of 8.231~days for the rotation period used in
our spot models since it more likely represents the rotation period
of a strong spot closer to the equator than the spot associated with the strongest period (assuming differential rotation in the same sense as the Sun).

Because the periodogram has power in the harmonics of the rotation period, we further investigated the presence of ellipsoidal variations.  We found that after removing the eclipse signature from the light curve, the harmonics also disappeared indicating that they are not associated with the periodic change in flux attributed to ellipsoidal variations.  This further supports our above conclusion that the change in flux from the presence of starspots is consistently great enough to mask the signature of the ellipsoidal variations.


\subsection{Spot Models for KOI-1003}

Light-curve inversion techniques can be employed to reconstruct stellar surfaces through the analysis of the shapes of the light curves.  Because the primary star is nearly two orders of magnitude more luminous than the secondary, we attribute the rotational modulation seen in the light curve to activity on the primary star.  As spots form and disappear, the modulations in the light curves change, often from one rotation cycle to the next \citep[e.g.,][]{roe11,roe13}.  Light-curve inversion methods use regularization procedures to determine a unique solution for each light curve.  Here, we use Light-curve Inversion (LI), a well-tested algorithm for surface reconstructions \citep{har00,roe11}.  LI breaks the stellar surface into a series of patches in bands of latitude where each patch is approximately the same area.  LI makes no \emph{a priori} assumptions of spot shape, number, or size, but takes as input the estimated photospheric and spot temperatures $T_\mathrm{phot}$ and $T_\mathrm{spot}$ and the angle of inclination $i$ of the rotation axis to the line of sight.  Limb darkening coefficients are also provided and are based upon the estimated $T_\mathrm{phot}$ and stellar parameters.  

From a compilation of starspot and photosphere temperatures from \citet{ber05}, we estimated the difference between the photosphere and the spot to be approximately $1300$ K, which gives $T_\mathrm{spot} \sim 3900$ K.  Changes in spot temperature will not affect the location of the starspots on the surface but may impact the overall size of the spot.  For limb darkening coefficients, we used the logarithmic coefficients provided by \citet{cla13}: $e = 0.7369$ and $f = 0.1359$.  Because we only have the \emph{Kepler} bandpass, the limb darkening is not a sufficient constraint on the starspots in latitude.  We emphasize that the spot latitudes in our reconstructions are not reliable on an absolute scale and are degenerate.  More advanced methods of imaging are required for a non-degenerate determination of spot latitude \citep[e.g., interferometric aperture synthesis imaging;][]{roe16}.

Although the system is eclipsing with an orbital angle of inclination of $i_\mathrm{orb} =86.0 \pm 0.5$, the angle of inclination of the stellar rotation axis is unknown.  We note that because $P_\mathrm{orb}$ and $P_\mathrm{rot}$ indicate that the system is nearly synchronized, we suggest the rotational angle of inclination is $i_\mathrm{rot} \approx 90^{\circ}$.  However, without a spectroscopic measurement of $v \sin i$, a measurement difficult to obtain for a star with $K_p = 16.209$ \citep{bro11}, we cannot know this value with certainty.  

Here, for the purposes of comparison to show consistency between the models, we considered five angles of inclination:  $i_\mathrm{rot} = 30^\circ, 45^\circ, 60^\circ, 75^\circ,$ and $90^\circ$.   We neglect $i_\mathrm{rot} = 0^\circ$ because this is a pole-on star and there will be no periodic, rotational modulations.   $i_\mathrm{rot} = 15^\circ$ is neglected because the resulting inversions were qualitatively very different from those for the other inclinations, since the small inclination leads to little modulation unless the spots are unrealistically large.  

As the light curve of KOI-1003 shows activity manifesting as both starspots and flares, we removed the data containing flares.   We identified the flares by eye \citep[as rapid increases in flux followed by an exponential decline, as described by][]{wal11} and manually removed the affected data.  Like the starspots, we assume that the flares are associated with the primary star due to its brightness compared to the secondary. 
We additionally removed the data for the primary transit and secondary eclipse.  The light curve free of systematic \emph{Kepler} variations (accounted for with the CBVs), transits, eclipses, and flares left only the features we believe to be the result of cool starspots.  We divided the light curves into sections the length of a single rotation period and binned the data into fifty bins of equal duration in order to reduce computation time ($\sim 8$ observations in each bin).

The light curves for individual rotation periods were inverted using LI.  The resultant surface reconstructions of these inversions are included in the appendix.  To analyze the starspots, we identified surface patches as being a starspot patch when the intensity of the patch is darker than 95\% of the average patch intensity.  We calculated the weighted average latitude and longitude.  We again note that the latitude information obtained from the inversions is not reliable, as only limited latitude information can be retrieved from a single photometric band.  Here, we only utilized the spot longitude.

\subsection{Persistent Starpots}

\begin{figure}
\includegraphics[angle=90,scale=.35]{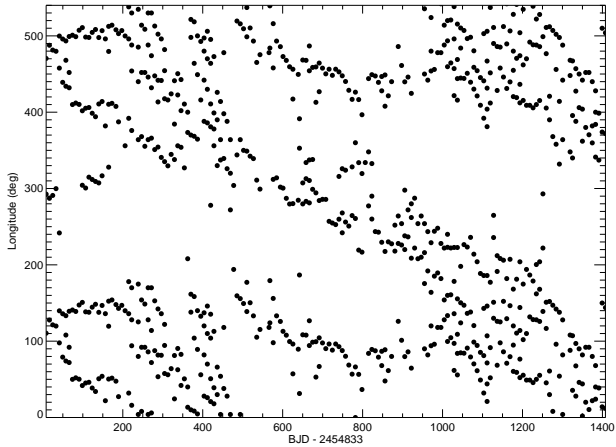}
\caption{Longitude of starspots of KOI-1003 plotted against time.  This sample plot is for the longitudes of the surface inversions for $i = 90^\circ$, which has two starspots with rotation periods that closely match the strongest two peaks in Figure \ref{pgram}.  The spot with the rotation period most closely matching that of the rotational period assigned to the system is found consistently around $100^\circ$ in longitude.  The other, higher latitude spot (assuming differential rotation in the solar sense given the negative slope present here) changes in longitude throughout the \emph{Kepler} light curve.
\label{longvtime}}
\end{figure}

To understand the motion of the starspots, we plotted starspot longitude against time (assigning the entire rotation the time of the first data point in the light curve), as in the example of Figure \ref{longvtime}.  For each inclination, we saw that there are two distinct spot structures (consisting of a number of spots that grow and disappear over time) that slowly change in longitude, suggesting that the starspots are rotating around the surface more slowly than the assumed stellar rotation.  We suggest that these regions of persistent starspots are ``active longitudes,'' as observed on other active stars \citep[e.g.,][and references therein]{ber05}.

To determine the rate of starspot rotation, we found the slope of the two regions where the motion of the spots over time are distinct, between BJD 2455318.64382 and 2455828.94975 (roughly between BJD $-$ 2454833 = 500 $-$1000 in Figure \ref{longvtime}).  The slopes ($^\circ/\mathrm{rotation \ period}$) and the associated rotation periods are found in Table \ref{slopeper}.  
The two prominent spot structures appear to have begun (in our observations) as closely-located, separated, and again approached each other over the course of the \emph{Kepler} light curve.  Because of their distinct slopes, the starspots appear to be located at different latitudes suggesting differential rotation.

These two spot structures are continuously present; therefore, we expected their rotation periods to be present in our period search.  Comparing the rotation periods predicted in Table \ref{slopeper} to those rotation periods detected in Figure \ref{pgram}, we found that the rotation periods are associated with the two strongest peaks.  We assume differential rotation in the same sense as the Sun and designate the spot with a shorter rotation period as the lower latitude spot (the spot consistently around a longitude of $100^\circ$ in Figure \ref{longvtime}).  We note that this low-latitude spot does not move across the surface in a linear way suggesting that the spot is changing latitude (see slight curvature between days 500 and 1000 (BJD $-$ 2454833) of the spot around longitude $= 100^\circ$ of Figure \ref{longvtime}).  Because of this, we expect a rotation period present deviating slightly from the low-latitude spot's rotation period.  The chosen rotation period (8.23~days) for the star is likely tied to the motion of this starspot.  The high-latitude spot has more linear trend suggesting that the rotation period of this starspot would be present in the periodogram.  The rotation period for the high-latitude spot reasonably matches the strongest peak present (8.30~days).

\begin{deluxetable*}{l c c c c}
\tabletypesize{\scriptsize}
\tablecaption{KOI-1003 Starspot Rotation Periods}
\tablewidth{0pt}
\tablehead{\colhead{Inclination ($^\circ$)} & \colhead{Low-latitude Spot} & \colhead{Period (day)} & \colhead{High-latitude Spot} &\colhead{Period (day)}\\
 & \colhead{Slope ($^\circ$/day)}  &  & \colhead{Slope ($^\circ$/day)}  & 
}
\startdata
$30$ & $-1.21 \pm 0.16$ & $8.259 \pm 0.004$ & $-2.42 \pm 0.21$ & $8.287 \pm 0.005$\\
$45$ & $-1.24 \pm 0.17$ & $8.260 \pm 0.004$ & $-2.59 \pm 0.22$ & $8.291 \pm 0.006$\\
$60$ & $-1.35 \pm 0.19$ & $8.262 \pm 0.005$ & $-2.67 \pm 0.23$ & $8.293 \pm 0.006$\\
$75$ & $-1.21 \pm 0.16$ & $8.259 \pm 0.004$ & $-2.38 \pm 0.24$ & $8.286 \pm 0.006$\\
$90$ & $-1.31 \pm 0.17$ & $8.261 \pm 0.004$ & $-2.56 \pm 0.23$ & $8.290 \pm 0.006$

\enddata
\label{slopeper}
\end{deluxetable*}\vspace{0.4cm}


\section{Discussion}

 In our analysis of KOI-1003, we have determined that the system is not a member of the 8~Gyr open cluster NGC 6791 \citep{cha99, moc02,moc05}.  Additionally, we show evidence that the primary star is not the main sequence star suggested by the \emph{Kepler} Input Catalog \citep{bro11}, \citet{pin12}, and \citet{hub14}.  We have determined the system to be an active RS CVn binary with a subgiant primary star and a main sequence secondary, the first such system to be identified using the \emph{Kepler} light curves.  By identifying and studying RS CVns in the \emph{Kepler} sample, we have access to unprecedented observations of rapidly-evolving stellar magnetism.  Long-term ground-based observations of RS CVns have led to studies of spot evolution and differential rotation, but the studies often depend on data that has been obtained over a few rotation periods \citep[e.g.,][]{roe11}.  With studies like this one, we see that spot evolution can occur over a single rotation period.  Our understanding of stellar magnetism benefits greatly from high-precision, high-cadence, longterm light curves like those obtained by the \emph{Kepler} satellite, as well as those that will be obtained by future missions including the Transiting Exoplanet Survey Satellite (TESS) and the CHaracterizing ExOPlanet Satellte (CHEOPS).  

The disposition of KOI-1003 has changed several times, flagging the system alternately as a potential planet-hosting star and an eclipsing binary.  The presence of starspots and their locations (with respect to the transit) impact the transit depth.  For KOI-1003, the effect is not to an extent that can definitively change the system's classification.  However, for systems with smaller companions or larger starspots, the ambiguity of the companion in this radius range is more troublesome.  
The mass-radius relation of exoplanets has been studied on
numerous occasions, with simple correlations described by
\citet{kan12} and \citet{wei14}. However, a size of $\sim 1 \ R_\mathrm{J}$ has considerable ambiguity as to the nature of the companion
without a mass measurement 
\citep[a $\sim 1 \ R_\mathrm{J}$ radius applies to a range of objects from giant planets through brown dwarfs to M dwarfs;][]{bar08,bar10}.   Without a mass estimate, the light curve alone cannot definitively be used to classify the nature of the companion.  Here, we have illustrated a way forward to obtaining realistic estimates of stellar parameters by isolating the eclipse signature from the starspots.  

We determined that the orbital period is well-matched with the rotation period of the primary star ($P_\mathrm{orb} - P_\mathrm{rot} \approx 0.13$ days, or $1.6\%$ of $P_\mathrm{orb}$) and that the system has an age of $3.0^{-0.5}_{+2.0}$~Gyr with the primary star evolved off of the main sequence.  The binary system is likely nearly evolved into a synchronized, circular orbit.  \citet{wal49} and \citet{zah89}, among others, have shown evidence that binary systems with $P_\mathrm{orb} \lesssim 10$ days will synchronize while on the main sequence, suggesting that $P_\mathrm{orb} = P_\mathrm{rot}$.  However, KOI-1003 is found here to have an orbital period that is slightly longer than the rotational period.  This discrepancy can be explained by noting that the spots used to determine the best estimate for the rotation period do not necessarily reflect the mean rotation rate of the star.  In order for this (near-)synchronization to have occurred for KOI-1003, the tidal forces of the primary and secondary would need to be significant \citep[see][and references therein]{maz08}, supporting our requirement of the secondary to be a main sequence star.    Since theory also determines that the orbit of KOI-1003 should be circular \citep[e.g.,][]{zah77}, we suggest that the system is a marginal case in the middle of this process, as its periods near the boundary of main-sequence circularization.  Therefore, because synchronization occurs before circularization \citep[e.g.,][]{maz08} and KOI-1003 is nearly synchronized, the binary will soon be circularized.

We found that the starspots modeled with LI, regardless of inclination, have spot rotation periods that match reasonably well those found in the two most significant peaks of the period search. 
We are unable to strongly constrain the stellar inclination without further spectroscopic investigations, but the near synchronization of the system suggests $i_\mathrm{rot} \approx 90^{\circ}$.  On the primary star of KOI-1003, we also determined the presence of long-lived active longitudes, where starspots appear to preferentially form.  These active longitudes have previously been inferred from observations of other stars \citep[e.g.,][and references therein]{ber05}, but have been shown, in at least one case, to be instead the photometric signal of ellipsoidal variations \citep[e.g.,][]{roe15b}.  We eliminated the possibility of ellipsoidal variations contributing to the active longitudes of KOI-1003 through our analysis.  

While improving an understanding of activity by observing spot evolution, this study also aimed to reveal the importance of understanding activity in the context of characterizing and detecting planets.  Recent works have emphasized that starspots are known to produce signals mistaken for planets \citep[e.g.,][]{kan16}.  
As a result, upcoming studies aimed at observing planets in the close-in habitable zones of low-mass stars (such as TESS and CHEOPS) will need to consider activity.  Our methods of analysis will be applied to a larger number of stars in future works such that the nature of the systems' spots and components may be realized.


\section*{Acknowledgements}

We thank S.\ T.\ Bryson, D.\ R.\ Ciardi, D.\ A.\ Caldwell, and B.\ T.\ Montet for their discussions on \emph{Kepler} and  KOI-1003 that improved the contents of this work. 
R.M.R.\ acknowledges support through the NASA Harriett G.\ Jenkins Pre-doctoral Fellowship Program.  Additional support for this project was provided through the Cycle 4 \emph{Kepler} Guest Observer Program (NASA grant NNX13AC17G) and NSF grant AST-1108963. 
This paper includes data collected by the \emph{Kepler} mission. Funding for the \emph{Kepler} mission is provided by the NASA Science Mission Directorate.  This research has made use of the NASA Exoplanet Archive, which is operated by the California Institute of Technology, under contract with the National Aeronautics and Space Administration under the Exoplanet Exploration Program.  This paper uses data from the United Kingdom Infrared Telescope (UKIRT), operated by the Joint Astronomy Centre on behalf of the Science and Technology Facilities Council of the U.K.


\section*{Appendix}
\begin{figure*}
\includegraphics[angle=0,scale=.85]{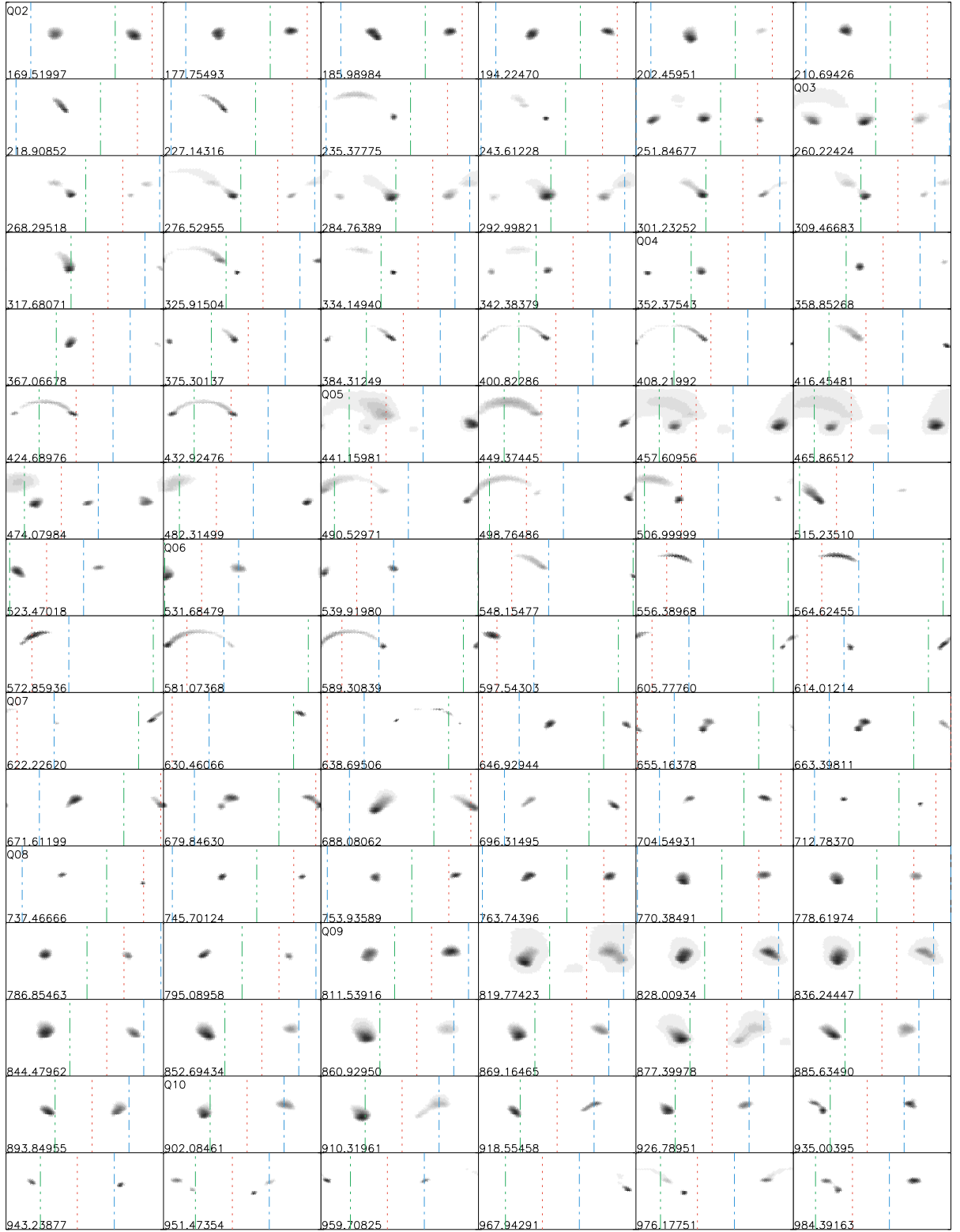}
\vspace{-1cm}
\caption{Panel of LI-reconstructed pseudo-Mercator surfaces of KOI-1003 for $i_\mathrm{rot} = 30^\circ$.  The beginning Barycentric Julian Date $-$ $2454833$ is included in the lower left corner of each plot.  Each plot maps longitude horizontally ($0-360^\circ$) and latitude vertically ($-90^\circ- + 90^\circ$).  As viewed by \emph{Kepler}, the center of the star at phase 0.0 is longitude $90^\circ$.  The star rotates such that the longitude decreases as time increase.  The blue single-dot-dashed line represents the point of the surface at which the companion crosses the surface as transit starts.  Assuming rotation and orbital motion are in the same direction, the green multiple-dot-dashed line represents the point on the surface at which the companion crosses as the transit ends and the red dashed line is the mid-point of the surface halfway through the transit.
\label{30A}}
\end{figure*}

\begin{figure*}
\includegraphics[angle=0,scale=.85]{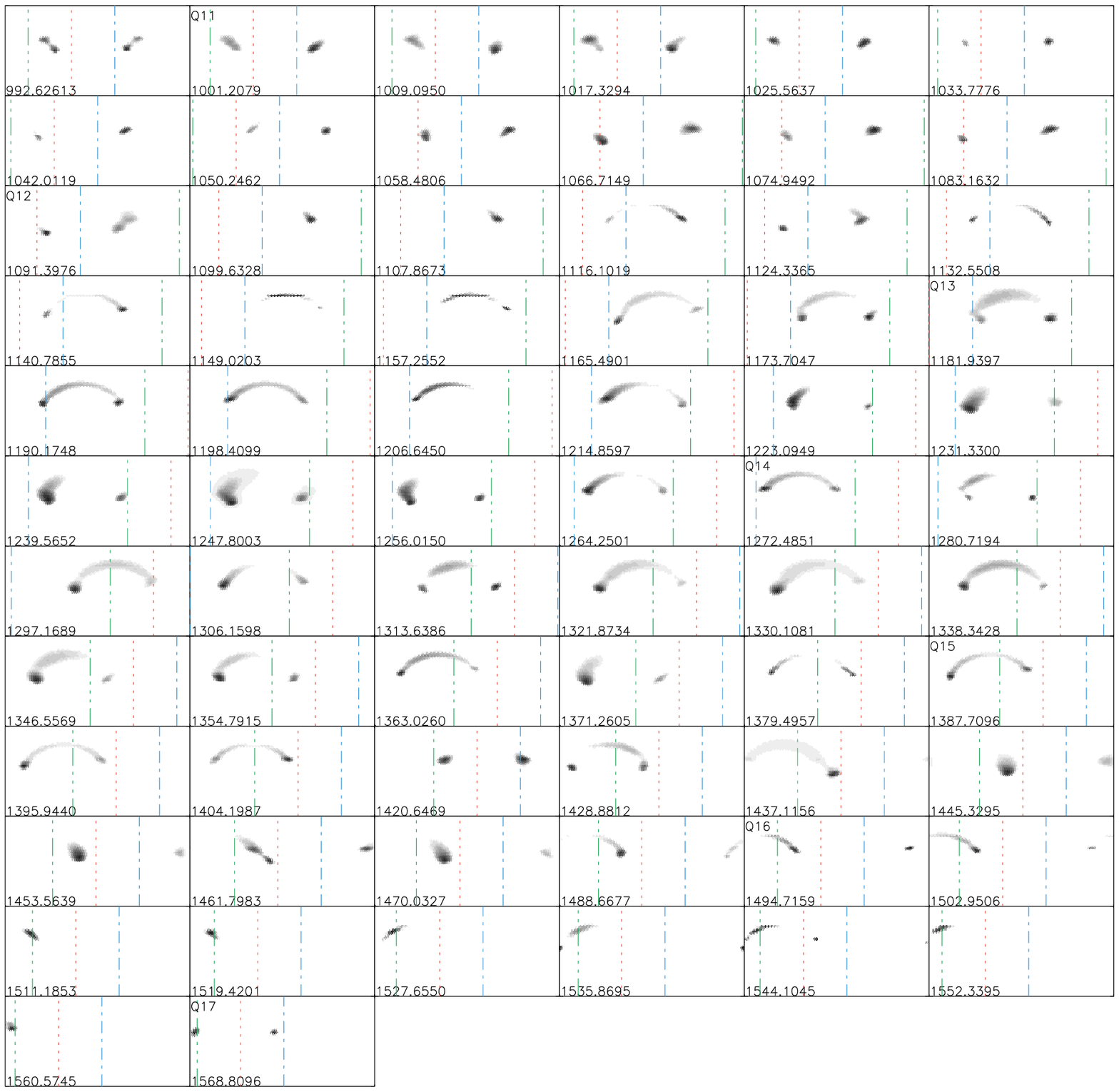}
\caption{Panel of LI-reconstructed pseudo-Mercator surfaces of KOI-1003 for $i_\mathrm{rot} = 30^\circ$, as in Figure \ref{30A}.
\label{30B}}
\end{figure*}

\begin{figure*}
\includegraphics[angle=0,scale=.85]{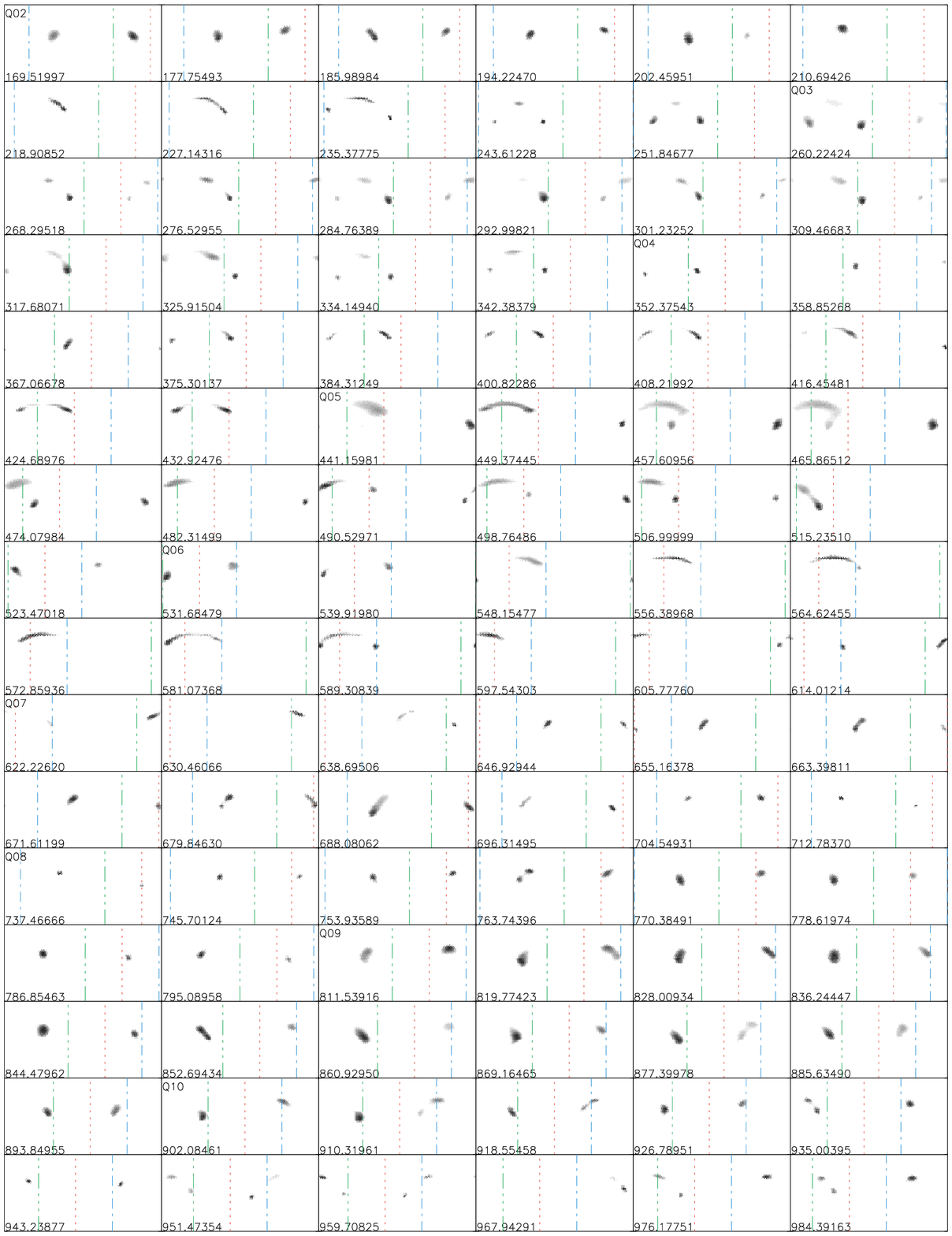}
\caption{Panel of LI-reconstructed pseudo-Mercator surfaces of KOI-1003 for $i_\mathrm{rot} = 45^\circ$, as in Figure \ref{30A}.
\label{45A}}
\end{figure*}

\begin{figure*}
\includegraphics[angle=0,scale=.85]{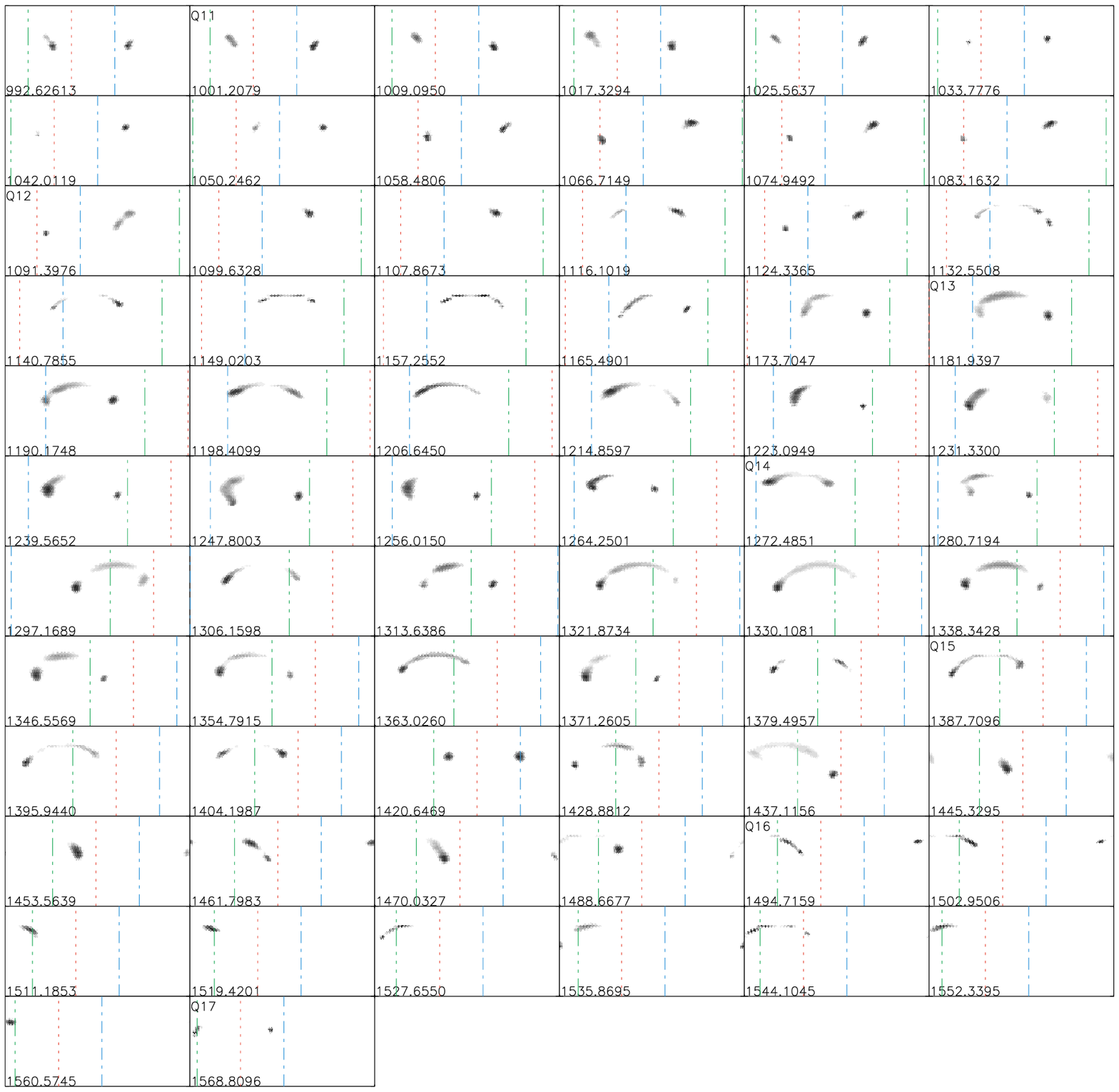}
\caption{Panel of LI-reconstructed pseudo-Mercator surfaces of KOI-1003 for $i_\mathrm{rot} = 45^\circ$, as in Figure \ref{30A}.
\label{45B}}
\end{figure*}

\begin{figure*}
\includegraphics[angle=0,scale=.85]{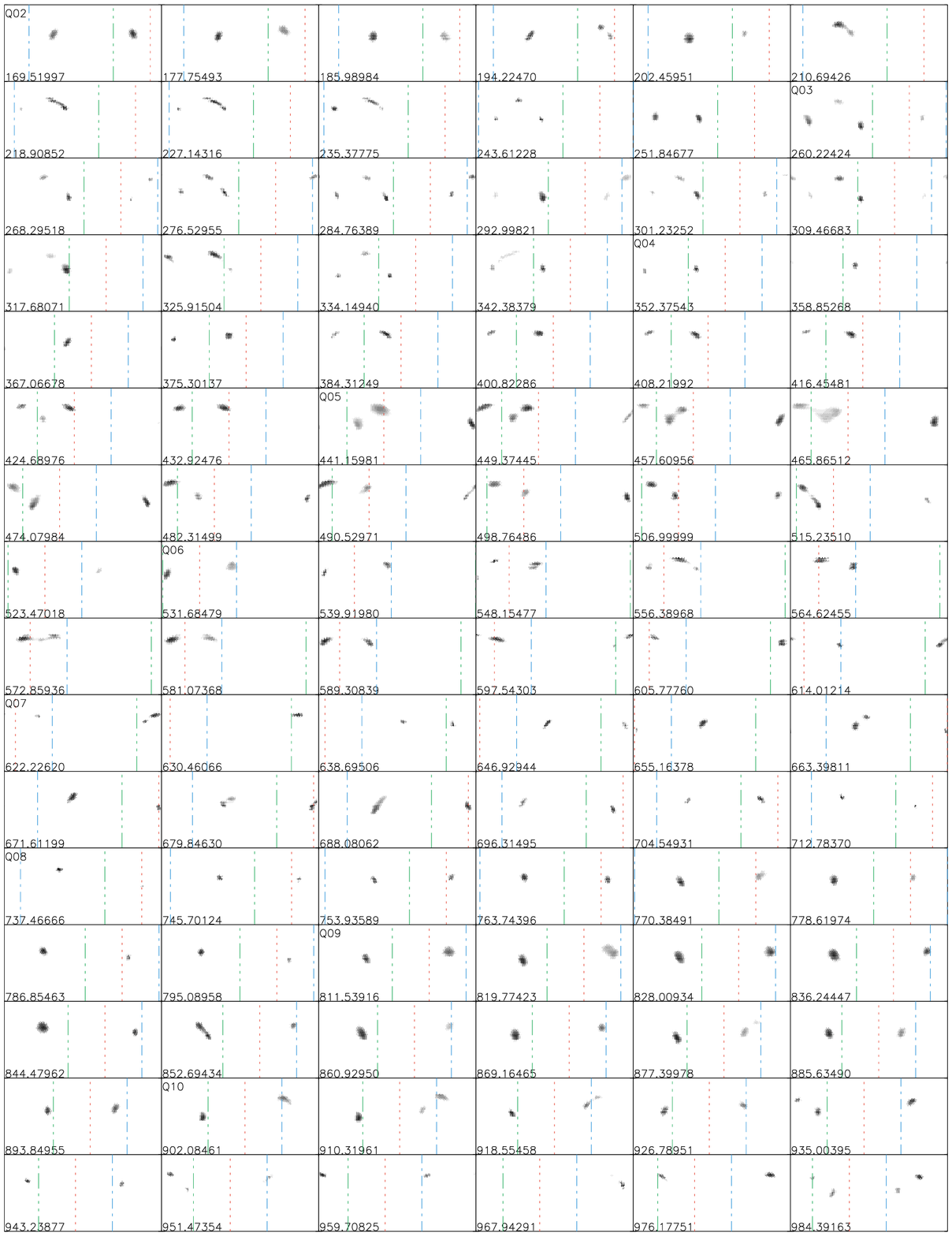}
\caption{Panel of LI-reconstructed pseudo-Mercator surfaces of KOI-1003 for $i_\mathrm{rot} = 60^\circ$, as in Figure \ref{30A}.
\label{60A}}
\end{figure*}

\begin{figure*}
\includegraphics[angle=0,scale=.85]{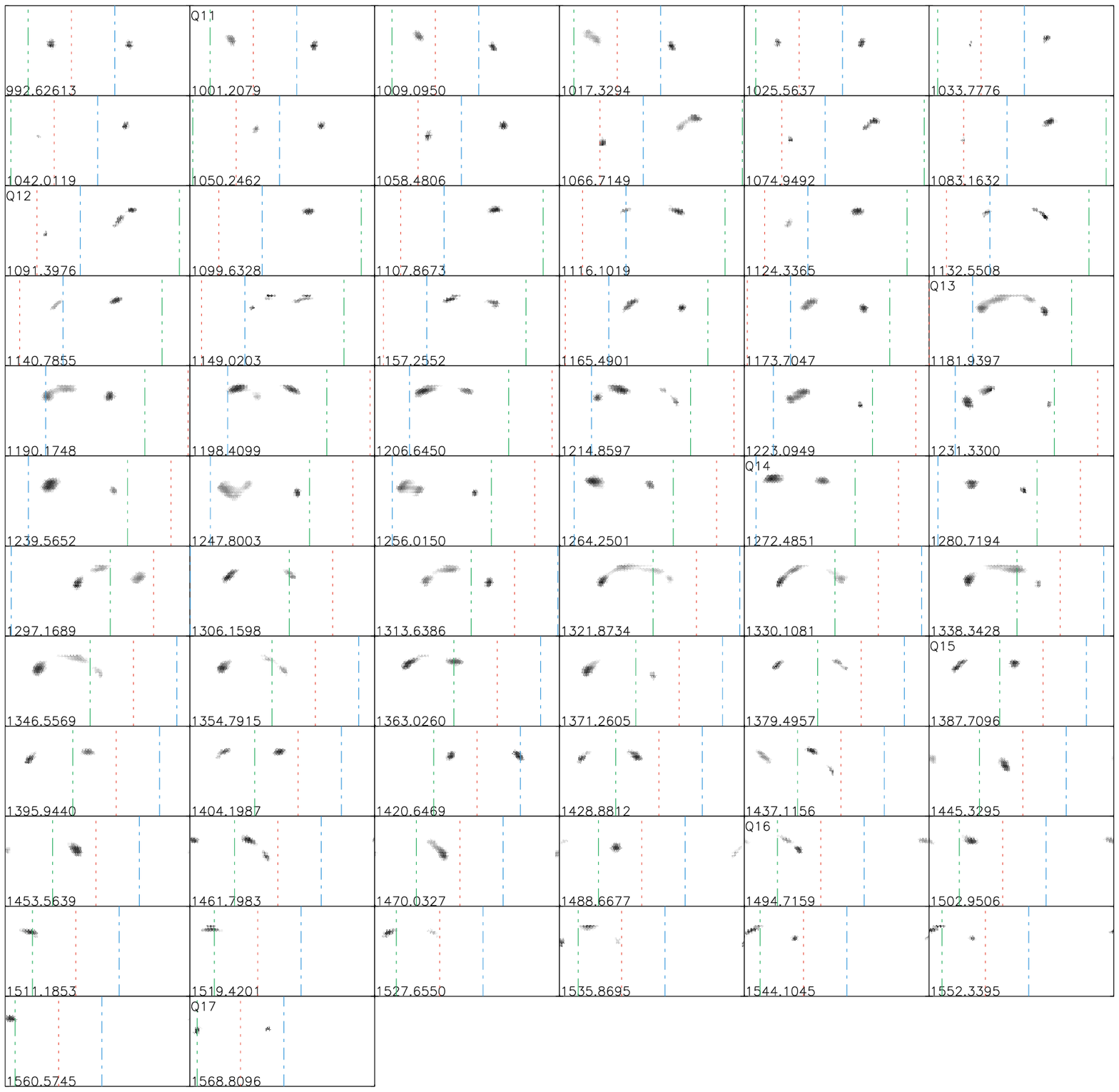}
\caption{Panel of LI-reconstructed pseudo-Mercator surfaces of KOI-1003 for $i_\mathrm{rot} = 60^\circ$, as in Figure \ref{30A}.
\label{60B}}
\end{figure*}

\begin{figure*}
\includegraphics[angle=0,scale=.85]{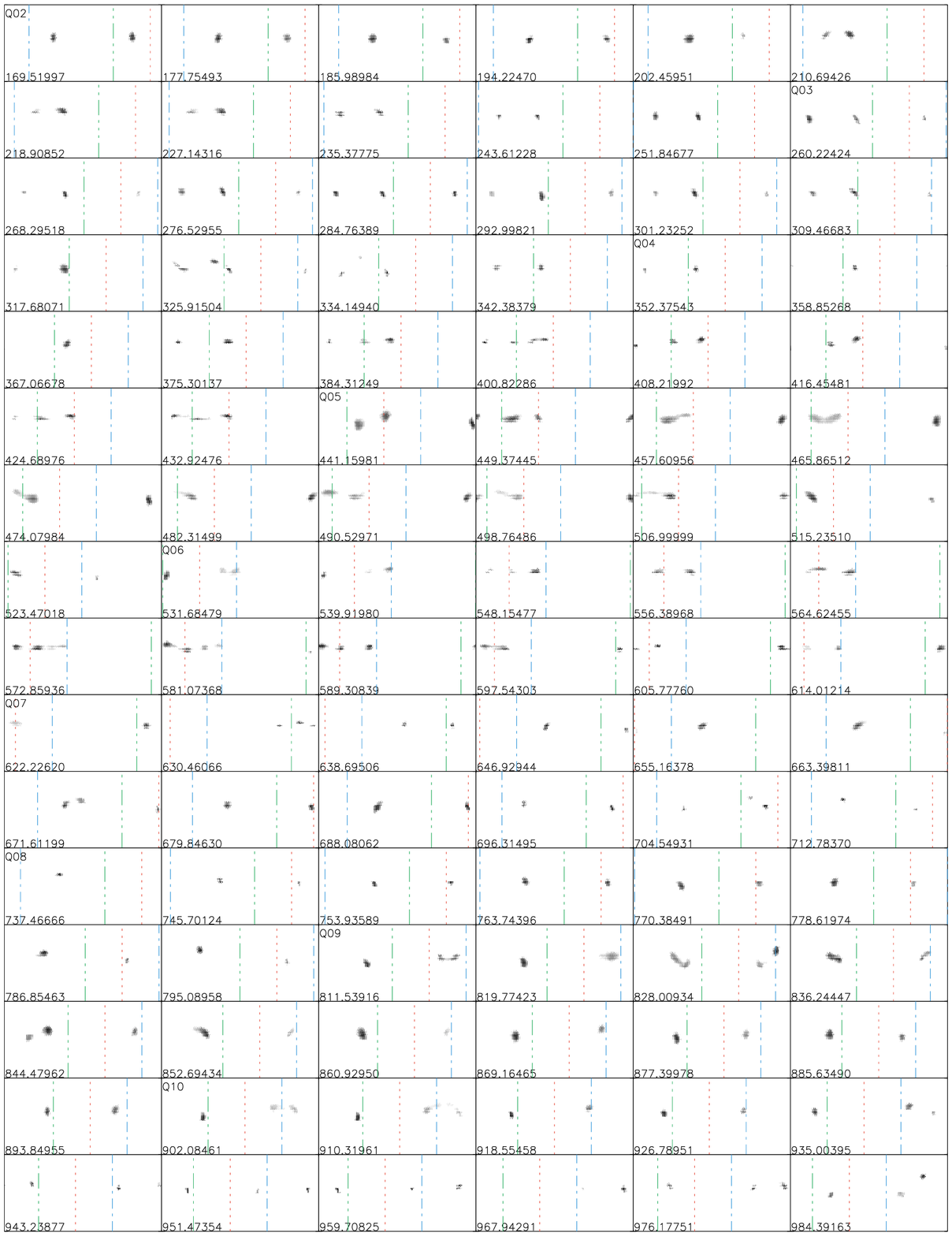}
\caption{Panel of LI-reconstructed pseudo-Mercator surfaces of KOI-1003 for $i_\mathrm{rot} = 75^\circ$, as in Figure \ref{30A}.
\label{75A}}
\end{figure*}

\begin{figure*}
\includegraphics[angle=0,scale=.85]{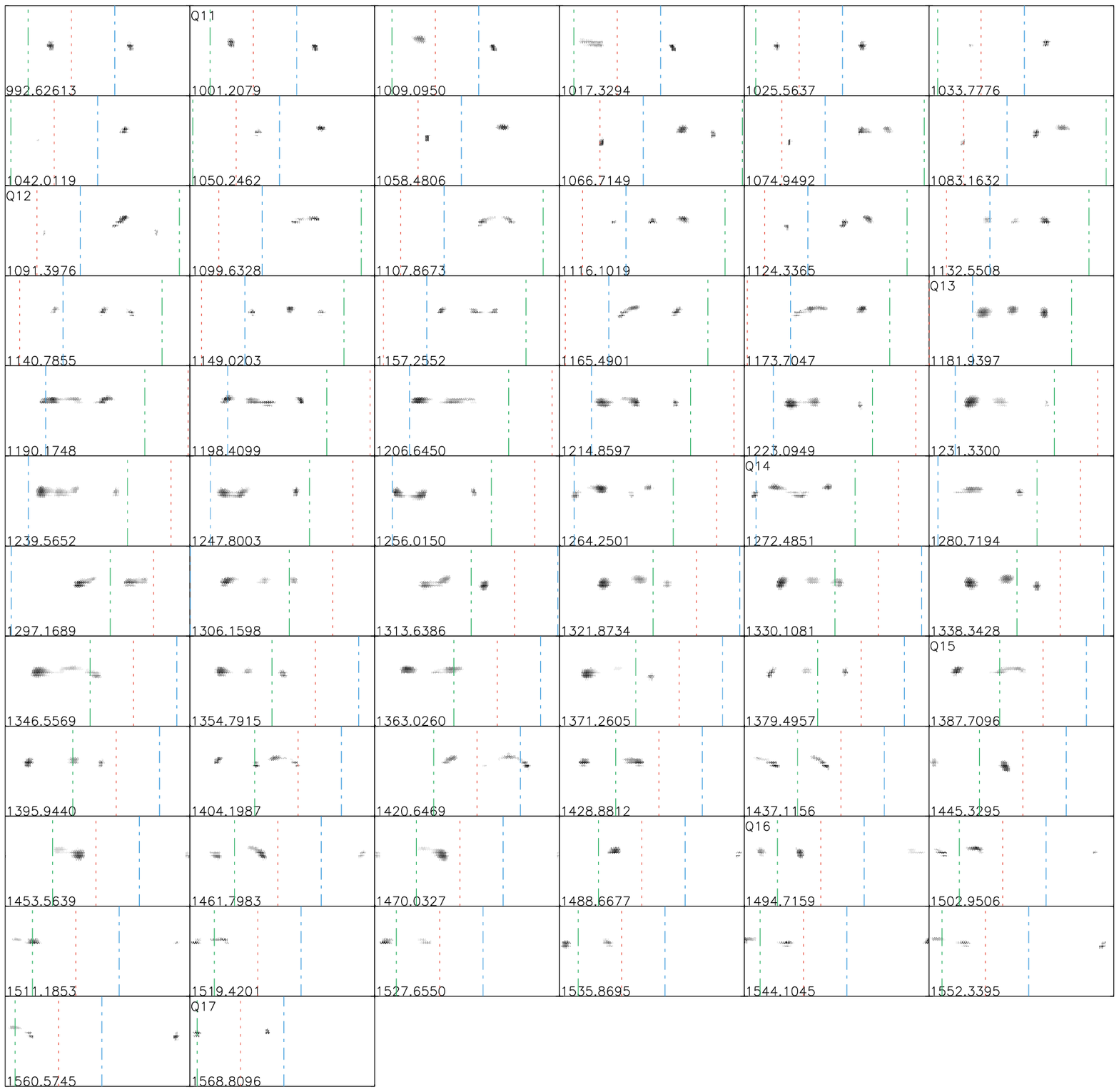}
\caption{Panel of LI-reconstructed pseudo-Mercator surfaces of KOI-1003 for $i_\mathrm{rot} = 75^\circ$, as in Figure \ref{30A}.
\label{75B}}
\end{figure*}

\begin{figure*}
\includegraphics[angle=0,scale=.85]{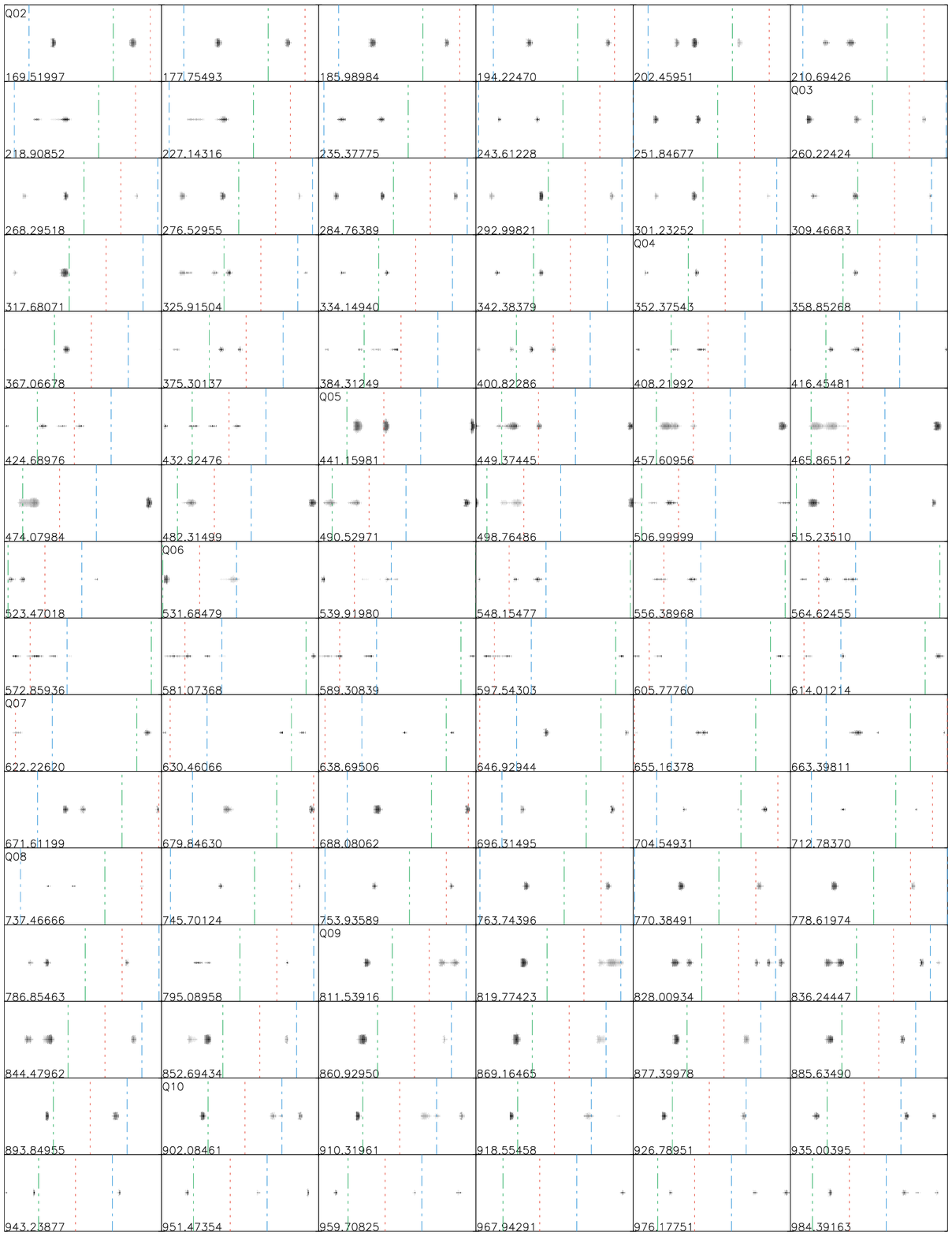}
\caption{Panel of LI-reconstructed pseudo-Mercator surfaces of KOI-1003 for $i_\mathrm{rot} = 90^\circ$, as in Figure \ref{30A}.
\label{90A}}
\end{figure*}

\begin{figure*}
\includegraphics[angle=0,scale=.85]{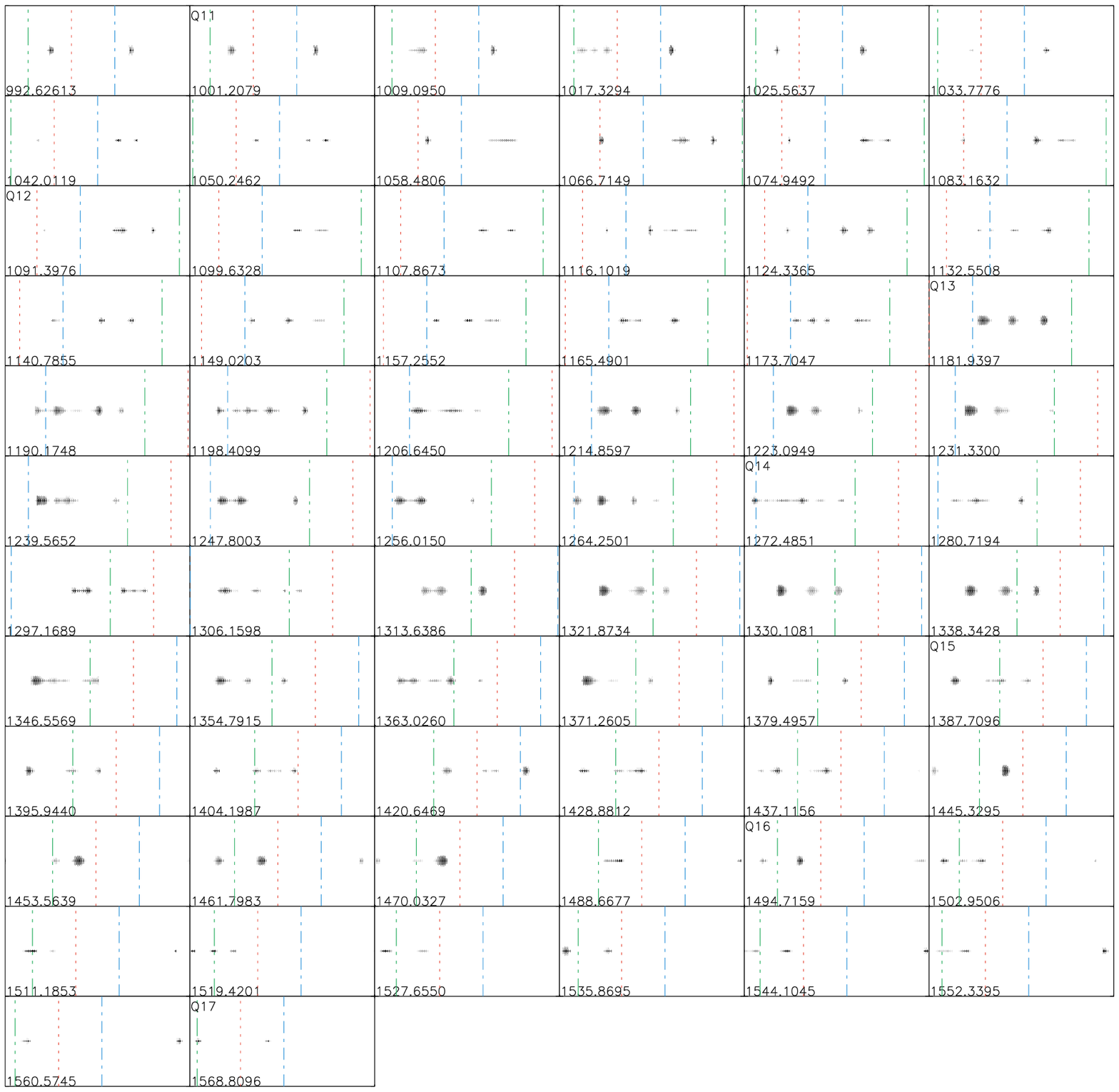}
\caption{Panel of LI-reconstructed pseudo-Mercator surfaces of KOI-1003 for $i_\mathrm{rot} = 90^\circ$, as in Figure \ref{30A}.
\label{90B}}
\end{figure*}

For five different inclination angles ($i_\mathrm{rot} = 30^\circ, 45^\circ, 60^\circ, 75^\circ,$ and $90^\circ$), we include the pseudo-Mercator maps of the results surface reconstructions from LI in Figures \ref{30A} - \ref{90B}.  Each spans $0 - 360^\circ$ in longitude horizontally and $-90^\circ - +90^\circ$ vertically.  The center of the star at phase 0.0, as seen by \emph{Kepler} is located at longitude $90^\circ$.  The left side of the plot, longitude $= 0^\circ$ is the limb of the star that has just rotated into view.  The convention of LI is that longitude \emph{decreases} as the star rotates.  We additionally include three lines on each surface plot, which represent the point on the surface where the companion first crosses the primary as the eclipse begins, the central point of the transit when the companion crosses the center of the primary, and the point on the surface where the companion last crosses the primary where the eclipse ends.

For each map, the long cadence \emph{Kepler} data were used with the CBVs removed from the SAP data.  The eclipses were removed, and the data were binned in fifty phase bins (to reduce computation time; $\sim 8$ observations in each bin).  Single-rotation period light curves were inverted with LI if the phase coverage was greater than $65\%$.  
As stated in Section 4, LI assumes the following input parameters:  $T_\mathrm{eff,1} = 5200$ K, $T_\mathrm{spot} = 3900$ K, and limb-darkening coefficients $e = 0.7369$ and $f = 0.1359$.

\begin{deluxetable*}{l c c c c c}
\tabletypesize{\scriptsize}
\tablecaption{rms Deviations between Observed and Reconstructed Light Curves (magnitudes)}
\tablewidth{0pt}
\tablehead{ \colhead{Angle of Rotational Inclination ($^\circ$)} & \colhead{Mean} & \colhead{Median} & \colhead{Minimum}  & \colhead{Maximum} & \colhead{Standard Deviation}
}
\startdata
$30$ & $0.0010$ & $0.0009$ & $0.0004$ & $0.0026$ & $0.0004$\\
$45$ & $0.0009$ & $0.0008$ & $0.0003$ & $0.0025$ & $0.0004$\\
$60$ & $0.0009$ & $0.0008$ & $0.0004$ & $0.0023$ & $0.0004$\\
$75$ & $0.0009$ & $0.0008$ & $0.0004$ & $0.0022$ & $0.0004$\\
$90$ & $0.0009$ & $0.0008$ & $0.0004$ & $0.0021$ & $0.0004$
\enddata
\label{rms}
\end{deluxetable*}

The maps presented were chosen from a series with varying rms values between the observed and reconstructed light curves.  The criteria for selecting a map is based upon identifying the amount of noise required to balance between fitting to the noise and smoothing the surface features.  Statistics on the rms values used for these reconstructions are found in Table \ref{rms}.  For a detailed discussion of LI, see \citet{har00} and \citet{roe11}, and its application to \emph{Kepler} data, see \citet{roe13}.


\end{document}